\documentclass[12pt]{article}
\usepackage[dvips]{graphicx}
\usepackage{amsmath}
\usepackage{amssymb}
\usepackage{graphicx}
\usepackage{color}
\usepackage{exscale,relsize}
\usepackage{multirow}

\renewcommand{\thefootnote}{\fnsymbol{footnote}}

\newcommand{\be}{\begin{equation}}
\newcommand{\ee}{\end{equation}}
\newcommand{\bea}{\begin{eqnarray}}
\newcommand{\eea}{\end{eqnarray}}
\newcommand{\beaa}{\begin{eqnarray*}}
\newcommand{\eeaa}{\end{eqnarray*}}

\newcommand{\mbeta}{{\mbox{\boldmath$\beta$}}}

\newcommand{\mb}{{\mbox{\boldmath$b$}}}
\newcommand{\mx}{{\mbox{\boldmath$x$}}}
\newcommand{\bmu}{{\mbox{\boldmath$\mu$}}}
\newcommand{\bomega}{{\mbox{\boldmath$\omega$}}}

\newcommand{\mX}{\mbox{$X$}}

\newcommand{\mw}{{\mbox{\boldmath$w$}}}

\newcommand{\mmu}{{\mbox{\boldmath$u$}}}
\newcommand{\my}{{\mbox{\boldmath$y$}}}
\newcommand{\me}{{\mbox{\boldmath$\varepsilon$}}}
\newcommand{\md}{{\mbox{\boldmath$d$}}}
\newcommand{\mdelta}{{\mbox{\boldmath$\delta$}}}
\newcommand{\mtheta}{{\mbox{\boldmath$\theta$}}}
\newcommand{\ba}{{\mbox{\boldmath$a$}}}

\newcommand{\mz}{{\mbox{\boldmath$z$}}}
\newcommand{\mee}{{\mbox{\boldmath$\epsilon$}}}

\newcommand{\mzero}{{\mbox{\boldmath$0$}}}

\newcommand{\ma}{\mathcal{A}}

\newcommand{\mI}{\mathcal{I}}
\newcommand{\mmm}{\mathcal{M}}

\newcommand{\mB}{\mathcal{B}}
\newcommand{\mXB}{\mbox{$X_{\mB}$}}

\newcommand{\SCAD}{{{scad}}}

\newcommand{\MCP}{{{mcp}}}

\long\def\symbolfootnote[#1]#2{\begingroup%
\def\thefootnote{\fnsymbol{footnote}}\footnote[#1]{#2}\endgroup}
\def\no{\noindent}

\begin{document}
\baselineskip= 24pt

\def\thetheorem{\arabic{section}.\arabic{theorem}}
\def\theremark{\arabic{remark}}

\begin{center}
\noindent{\Large A Novel Approach for Fast Detection of Multiple
Change Points in Linear Models}

\noindent{ Xiaoping Shi$^{\mbox{{\small a}}}$, Yuehua Wu$^{\mbox{{\small a}}}$ and Baisuo
Jin$^{\mbox{{\small b}}}$}

\bigskip
\noindent{${}^a$Department of Mathematics and Statistics, York University, Toronto, Ontario, Canada;
${}^b$Department of Statistics and Finance, University of Science and Technology of China, Hefei,
Anhui, China}
\end{center}

\bigskip

\noindent{\bf Abstract} A change point problem occurs in many
statistical applications. If there exist change points in a model,
it is harmful to make a statistical analysis without any
consideration of the existence of the change points and the results
derived from such an analysis may be misleading. There are rich
literatures on change point detection. Although many methods have
been proposed for detecting multiple change points, using these
methods to find multiple change points in a large sample seems not
feasible. In this article, a connection between multiple change
point detection and variable selection through a proper segmentation
of data sequence is established, and a novel approach is proposed to
tackle multiple change point detection problem via the following two
key steps: (1) apply the recent advances in consistent variable
selection methods such as SCAD, adaptive LASSO and MCP to detect
change points; (2) employ a refine procedure to improve the accuracy
of change point estimation. Five algorithms are hence proposed,
which can detect change points with much less time and more accuracy
compared to those in literature. In addition, an optimal
segmentation algorithm based on residual sum of squares is given.
Our simulation study shows that the proposed algorithms are
computationally efficient with improved change point estimation
accuracy. The new approach is readily generalized to detect multiple
change points in other models such as generalized linear models and
nonparametric models.


\setcounter{equation}{0}

\vspace{.3in} \noindent{\textbf{1. Introduction}}

\noindent The most popular statistical model used in practice is a linear model, which has been
extensively studied in the literature. This model is simple and can be used to approximate a nonlinear
function locally. However there may be change points in a linear model such that the regression
parameters may change at these points. Thus if there do exist change points in a linear model, the
linear model is actually a segmented linear model.

A change point problem occurs in many statistical applications in
the areas including medical and health sciences, life science,
meteorology, engineering, financial econometrics and risk
management. To detect all change points are of great importance in
statistical applications. If there exists a change point, it is
harmful to make a statistical analysis without any consideration of
the existence of this change point and the results derived from such
an analysis may be misleading. There are rich literatures on change
point detection, see, e.g., Cs\"{o}rg\H{o} and Horv\'{a}th (1997)
and Chen and Gupta (2000).

Compared with the detection of one change point, to locate all
change points is a very challenge problem. Although, it has been
studied in literature (see Davis, Lee, and Rodriguez-Yam (2006), Pan
and Chen (2006), and Kim, Yu and Feuer (2009), and Loschi, Pontel
and Cruz (2010) among others), a powerful and efficient method still
needs to be explored. Thus this paper is mainly concerned with the
multiple change point detection problem in linear regression.

Consider a linear model with $K_0\leqslant K_U<\infty$ multiple
change points located at $a^{(0)}_{1,n},\ldots, a^{(0)}_{K_0,n}$:
\begin{eqnarray}
y_{i,n}&=&\sum_{j=1}^q
x_{i,j,n}\beta_{j,0}+\sum_{\ell=1}^{K_0}\sum_{j=1}^q
x_{i,j,n}\delta_{j,0}^{(\ell)}I(a^{(0)}_{\ell,n}<
i\leqslant n)+\varepsilon_{i,n}\nonumber\\
&=&\mx_{i,n}^T\left[\mbeta_0+\sum_{\ell=1}^{K_0}\mdelta_{\ell,0}
I(a^{(0)}_{\ell,n}< i\leqslant n)\right]+\varepsilon_{i,n},\quad
i=1,\ldots, n,\label{cp}
\end{eqnarray}
where $\{\mx_{i,n}=(x_{i,1,n},\ldots,x_{i,q,n})^T\}$ is a sequence
of $q$-dimensional predictors, $\mbeta_0=(\beta_{1,0},\ldots,
\beta_{q,0})^T$ $\not=$ ${\bf 0}$ is unknown $q$-dimensional vector
of regression coefficients, $K_0$ is unknown number of change
points, $a^{(0)}_{1,n}$, $\ldots$, and $a^{(0)}_{K_0,n}$ are unknown
change point locations (or change points), $\mdelta_{\ell,0}$,
$1\leqslant \ell\leqslant K_0$, denote unknown amounts of changes in
regression coefficient vectors at change points, and
$\varepsilon_{1,n},\ldots,\varepsilon_{n,n}$ are random errors. In
this paper, we assume that $K_U$ is an upper bound of $K_0$. Set
$a^{(0)}_{K_0+1,n}=n$. If there is no change point, $K_0=0$ and the
model (\ref{cp}) becomes
\[y_{i,n}=\sum_{j=1}^q x_{i,j,n}\beta_{j,0}+\varepsilon_{i,n},\quad i=1,\ldots, n.\]
Otherwise, $K_0\geqslant 1$, and we assume that \be
0<a^{(0)}_{\ell,n}/n\rightarrow \tau_\ell<1, \quad\mbox{for
$1\leqslant \ell\leqslant K_0$}.\label{cdcp1}\ee If $K_0\geqslant2$,
we assume that \be\min_{1\leqslant\ell\leqslant
K_0-1}(\tau_{\ell+1}-\tau_\ell)>0\label{cdcp2}\ee is unknown. The
problem studied in this paper is to estimate $K_0$, $a^{(0)}_{1,n}$,
$\ldots$, and $a^{(0)}_{K_0,n}$ or in other words to detect multiple
change points. If there is no confusion, the superscript ``{(0)}'',
subscript ``0'', and subscript $n$ will be suppressed.

For detecting multiple change points, it may be convenient to
consider the following linear model with probable multiple change
points located at $1<a_{1,n}<\cdots< a_{K,n}<n$
\begin{eqnarray}
y_i&=&\mx_i^T\left[\mbeta+\sum_{\ell=1}^{K}\mdelta_{\ell}
I(a_{\ell,n}< i\leqslant n)\right]+\varepsilon_i,\quad i=1,\ldots,
n,\label{cp1}
\end{eqnarray}
where $\mbeta$, $\mdelta_{1}$, $\ldots$, $\mdelta_{K}$ are unknown
$q$-dimensional parameter vectors. We can instead test the following
null hypothesis: \beaa H_0:&&\mbox{There is no change point, i.e.,
for
any $1< a_{1,n}<\cdots<a_{K,n}<n$},\\
&&\mbox{$\mdelta_\ell=(\delta_{1}^{(\ell)}$, $\ldots$,
$\delta_{q}^{(\ell)})^T$}\mbox{ $=$ ${\bf 0}$ for any $\ell\in
\{1,\ldots,K\}$},\ \mbox{where }1\leqslant K\leqslant K_U\eeaa
versus the alternative hypothesis: \beaa H_1: &&\mbox{There exist
$1\leqslant K\leqslant K_U$ change points, i.e., there exist $1<
a_{1,n}<\cdots<a_{K,n}<n$}\\&&\mbox{such that
$\mdelta_\ell=(\delta_{1}^{(\ell)},\ldots,\delta_{q}^{(\ell)})^T\neq\mzero$}\
\mbox{for any $\ell\in \{1,\ldots,K\}$.}\eeaa Many classical methods
have been given in literature for detecting change points, which
include the popular model selection based change point detection
method and the well known cumulative sum (CUSUM)  method. However
the amounts of computing time required by these two typical change
point detection methods are respectively $O(2^n)$ and $O(n^2)$. When
$n$ is very large, using these methods to find multiple change
points seems not feasible.

If the set of all true change points in the model (\ref{cp1}) is a
subset of $\{a_{\ell,n}, \ 1\leqslant \ell \leqslant K\}$, it is
easy to see that $a_{j,n}$ is a change point if and only if
$\mdelta_j\not={\bf 0}$. We rewrite (\ref{cp1}) as follows:
\begin{eqnarray}
\my_n=\mX_n \tilde{\mbeta}+\me_n,\label{reg0}
\end{eqnarray}
where $\my=(y_1,y_2,\cdots,y_n)^T$,
$\tilde{\mbeta}=({\mbeta^T},\mdelta_1^T,\ldots,\mdelta_{{K}}^T)^T$,
$\me_n=(\varepsilon_1,\varepsilon_2,\ldots,\varepsilon_n)^T$, and
\begin{eqnarray*}
\mX_n&=&\left( \begin{array}{ccccc} X_{(0,1)} &0_{(0,1)}&0_{(0,1)}& \cdots & 0_{(0,1)}\\
X_{(1,2)} &X_{(1,2)}&0_{(1,2)}& \cdots & 0_{(1,2)}\\
\vdots &\vdots &\ldots &\ldots &\vdots \\
X_{(K,K+1)} &X_{(K,K+1)}&X_{(K,K+1)}& \cdots & X_{(K,K+1)}\\
\end{array} \right)_{n\times (K+1)q}\end{eqnarray*}
with $0_{(j-1,j)}$ is a zero matrix of dimension
$(a_{j,n}-a_{j-1,n})\times q$, and $a_{0,n}=0$,
\begin{eqnarray}X_{(j-1,j)}=\left( \begin{array}{ccc}
x_{a_{j-1,n}+1,1}& \cdots & x_{a_{j-1,n}+1,q}\\
\vdots &\cdots &\vdots \\
x_{a_{j,n},1}& \cdots & x_{a_{j,n},q}
\end{array} \right)_{(a_{j,n}-a_{j-1,n})\times q}\quad \mbox{for $j=1,\ldots,K+1$.}\nonumber\end{eqnarray}
Thus to detect all the true change points and remove the pseudo
change points in (\ref{cp1}) can be considered as a variable
selection problem for the linear regression model (\ref{reg0}), and
we may tackle the problem by employing variable selection methods.
This leads us to explore a possibility by first properly segmenting
data sequence and then applying variable selection methods and/or
other methods for detecting probable multiple change points.

The paper is arranged as follows. The segmentation of data sequence
and multiple change point estimation are discussed in Section 2.
Five algorithms for detecting probable multiple change points are
proposed in Section 3. Simulation studies and practical
recommendations are given in Section 4. Two real data examples are
provided in Section 5.

Throughout the rest of the paper, ${\bf 1}_q=(1,\ldots,1)^T$ is the
$q$-dimensional vector, $I_q$ is the $q\times q$ identity matrix, an
indicator function is written as $I(\cdot)$, the transpose of a
matrix $A$ is denoted by $A^T$, and $\lfloor c\rfloor$ is the
integer part of a real number $c$. For a vector $\ba$, $\ba^T$ is
its transpose, $\ba(j)$ is its $j$th component, $|\ba|$,  $\|\ba\|$
and $\|\ba\|_\infty$ are respectively its $L_1$-norm, $L_2$-norm
(Euclidean norm) and $L_\infty$ norm. If $\ma$ is a set, its
complement and its size are denoted by $\bar{\ma}$ and $|\ma|$,
respectively. In addition, the notations ``$\rightarrow_p$'' and
``$\rightarrow_d$'' denote convergence in probability and
convergence in distribution, respectively. Furthermore, the
$(1-\alpha)$th quantile of the chi-square distribution with $\ell$
degrees of freedom is denoted by $\chi^2_{\alpha,\ell}$.

\vspace{.2in} \noindent{\textbf{ 2. Segmentation and Change Point Estimation}}

\noindent For a multiple change point detection problem, the multiple change point locations are
unknown and in practice their approximate locations within a permissible range is main concern, which
inspires us to partition the data sequence to search for change points. We thus divide the data
sequence into $p_n+1$ segments. Let $m=m_n=\lfloor n/(p_n+1)\rfloor$. The segmentation is such that
the first segment has length $0<m\leqslant n-p_nm\leqslant c_0m$ with some $c_0\geqslant1$ and each of
the rest $p_n$ segments has length $m$. Without loss of generality, we assume that $p_n\to\infty$ as
$n\to\infty$. The partition of the data sequence yields the following segmented regression model:
\begin{eqnarray}
y_i&=&\mx_i^T\left[\mbeta+\sum_{\ell=1}^{p_n}\md_{\ell}
I\left(n-(p_n-\ell+1)m< i\leqslant
n\right)\right.\nonumber\\&+&\left.\sum_{\ell=1}^{p_n}\bomega_{\ell}(i)
I\left(n-(p_n-\ell+1)m< i\leqslant
n-(p_n-\ell)m\right)\right]+\varepsilon_i,\quad i=1,\ldots,
n,\label{newcp}
\end{eqnarray}
where two sets $\{\md_1,\ldots,\md_{p_n}\}$ and
$\{\mzero,\mdelta_1,\ldots,\mdelta_{K_0}\}$ are equal, and
$\{\bomega_\ell\}$ are defines as follows: if there is a change
point located in $\{n-(p_n-\ell+1)m+1,\ldots, n-(p_n-\ell)m-1\}$,
say $a_{k,n}$, then
\[\bomega_\ell(i)=\left\{\begin{array}{ll}
-\mdelta_k,& n-(p_n-\ell+1)m<i\leqslant a_{k,n}<n-(p_n-\ell)m,\\
\mzero,& \mbox{elsewhere};\end{array}\right.\] otherwise,
\[\bomega_\ell(i)=\mzero, \quad i=1,\ldots,n.\] The model (\ref{newcp}) can be written as
\begin{eqnarray}
\my_n=\tilde{X}_n
\mtheta_n+X_\omega\sum_{\ell=1}^{p_n}\vec{\bomega}_\ell+\me_n,\label{reg}
\end{eqnarray}
where $\my_n$ and $\me_n$ are defined in Section 1,
$\mtheta_n=(\theta_1,\ldots,
\theta_{q(p_n+1)})^T=({\mbeta^T},\md_1^T,\ldots,\md_{{p_n}}^T)^T$,
$\md_r=(d_{r1}, \ldots,d_{rq})^T$, $r=1,\ldots,{p_n}$,
\begin{eqnarray}
\tilde{X}_n&=&\left( \begin{array}{ccccc} X_{(1)} &0_{m
\times q}&0_{m\times q}& \cdots & 0_{m\times q}\\
X_{(2)} &X_{(2)}&0_{m\times q}& \cdots & 0_{m\times q}\\
\vdots &\vdots &\ldots &\ldots &\vdots \\
X_{(p_n+1)} &X_{(p_n+1)}&X_{(p_n+1)}& \cdots & X_{(p_n+1)}\\
\end{array} \right)_{n\times (p_n+1)q}=(X_n^{(1)},\ldots,X_n^{(p_n+1)} )\label{xx}\end{eqnarray}
with ${X}_n^{(j)}=(\mzero_{q\times m},\ldots,\mzero_{q\times m},
X_{(j)}^T,\ldots,X_{(p_n+1)}^T)^T$, \beaa\mX_{(1)}&=&\left(
\begin{array}{ccc}
x_{1,1}& \cdots & x_{1,q}\\
\vdots &\cdots &\vdots \\
x_{n-p_nm,1}& \cdots & x_{n-p_nm,q}
\end{array} \right)_{(n-p_nm)\times q},\eeaa
\beaa \mX_{(j)}&=&\left( \begin{array}{ccc}
x_{n-(p_n-j+2)m+1,1}& \cdots & x_{n-(p_n-j+2)m+1,q}\\
\vdots &\cdots &\vdots \\
x_{n-(p_n-j+1)m,1}& \cdots & x_{n-(p_n-j+1)m,q}
\end{array} \right)_{m\times q},\quad \mbox{for $j=2,\ldots,
p_n+1$,}\nonumber\eeaa $X_\omega=\text{diag}(\mx_1^T,\ldots,
\mx_n^T)$, and
$\vec{\bomega}_\ell=(\bomega_\ell^T(1),\ldots,\bomega_\ell^T(n))^T$.
It is easy to see that $\mx_\omega\equiv
X_\omega\sum_{\ell=1}^{p_n}\vec{\bomega}_\ell$ is an $n$ dimensional
vector and all its elements excluding at most $K_0(m-1)$ of them are
zeros. It is noted that in Harchaoui and Levy-Leduc (2008), the
mean-shift model is considered and the length of each of their
segments is only 1.

Consider a special case that each true change point is at an end of
a segment. Then an end of a segment is a true change point if and
only if the corresponding $\md_r\not={\bf 0}$. Thus to locate all
the true change points in (\ref{cp}) is equivalent to carry out
variable selection. Since $p_n\to\infty$, we may take advantage of
the recent advances in consistent variable selection methods for a
linear regression model as (\ref{reg}) with a large number of
regression coefficients, which include the SCAD (Fan and Li (2001)),
the adaptive LASSO (Zhou (2006)), and the MCP (Zhang (2010)) among
others.

Let us examine the relationship between the models (\ref{cp}) and
(\ref{reg}). It can be seen that under the null hypothesis $H_0$,
$\mbeta=\mbeta_0$, and $\md_r=\mzero$, $r\in\{1,\cdots,p_n\}$. We
now assume that $H_1$ hold. Thus, there exist $\{r_k,\
k=1,\cdots,K_0\}$ such that $a_{k,n}\in \{n-p_nm+(r_k-1)m,\ldots,
n-p_nm+r_km-1\}$. Since $K_0$ is finite with an upper bound $K_U$,
in view of (\ref{cdcp1}) and (\ref{cdcp2}), it follows that
\be\mbeta=\mbeta_0,\quad
\md_{r_k-1}=\mzero,\quad\md_{r_k}=\mdelta_k\not=\mzero,\quad\mbox{
and $\md_{r_k+1}=\mzero$}\label{hh1}\ee for large $n$. Thus in order
to detect all the change points $\{a_{1,n},\ldots,a_{K_0,n}\}$, we
may estimate $\{\md_{i}\}$ in advance.

The following assumptions are made for investigating the asymptotic
properties of the estimates of $\{{\md}_{i}\}$:

\noindent{\bf Assumption C1.}\quad
$\sum_{i=s}^t\mx_i\mx_i^T/(t-s)\rightarrow W>0$ as
$t-s\rightarrow\infty.$

It is noted that Assumption C1 is a common assumption made in change
point analysis for a mean shift model. Under Assumption C1, it can
be shown that $\mX_{(1)}^T\mX_{(1)}/(n-p_nm)\rightarrow W>0$, and
$\mX_{(i)}^T\mX_{(i)}/m\rightarrow W>0$ for
$i\in\{2,\ldots,p_n+1\}$.

\emph{Remark 1.} \quad Assumption C1 is similar to Condition (b) in
Zhou (2006). If we only consider the consistency of change point
estimators, Assumption C1 can be relaxed to the following weaker
one: For $b_1,b_2>0$,
$b_1I_q\leqslant\sum_{i=s}^t\mx_i\mx_i^T/(t-s)\leqslant b_2I_q$ when
$t-s$ is large enough.

\noindent{\bf Assumption C2.}\quad $\{\varepsilon_i,\quad
i=1,2,\ldots\}$ is a sequence of independently and identically
distributed (i.i.d.) random variables with mean 0 and variance
$\sigma^2$.

\emph{Remark 2.} \quad This assumption can be replaced by a weaker
assumption of the strong mixing condition in (2.1) in Kuelbs and
Philipp (1980), which adapts to the autoregressive models in Davis,
Huang and Yao (1995) and Wang, Li and Tsai (2007). Let
$\{\varepsilon_i,\ i=1,2,\ldots\}$ be a weak sense stationary
sequence of random variables with mean 0 and $(2+\delta)$th moments
for $0<\delta\leqslant 1$ that are uniformly bounded by some
positive constant. Suppose that $\{\varepsilon_i,\ i=1,2,\ldots\}$
satisfies the strong mixing condition $|P(AB)-P(A)P(B)|\leqslant
\rho(n)\downarrow 0$ for all $n$, $s\geqslant 1$, all $A\in
\mmm_1^s$ and $B\in\mmm_{s+n}^\infty$, where $\mmm_a^b$ is the
$\sigma$-field generated by the random vectors
$\varepsilon_a,\varepsilon_{a+1},\cdots,\varepsilon_b$, and
$\rho(n)<<n^{-(1+t)(1+2/\delta)}$ for some $t>0$. Then Theorem 4 and
Lemma 3.4 in Kuelbs and Philipp (1980) warrant the same results as
given in Theorems 1-3 below.

For simple presentation below, we assume that each of
$\{\mX_{(r)}\}$ is of full rank in this paper. If a $\mX_{(r)}$ is
not of full rank, Moore-Penrose matrix inverse can be used instead
of the matrix inverse.

\vspace{.2in} \noindent{\textsf{2.1. Estimate $\{\md_{i}\}$ by least squares}}

\noindent By least squares method, we estimate $\md_r$, $r=1,\ldots,{p_n}$, as follows:
\be\hat{\md}_{r}=\left(\mX^T_{(r+1)}\mX_{(r+1)}\right)^{-1}\mX^T_{(r+1)}\my^{(r+1)}-
\left(\mX^T_{(r)}\mX_{(r)}\right)^{-1}\mX^T_{(r)}\my^{(r)},\quad r=1,\ldots,p_n,\label{ddd}\ee where
$\my^{(1)}=(y_1,\ldots,y_{n-p_nm})^T$, and $\my^{(r)}=(y_{n-(p_n-r+2)m+1},\ldots,y_{n-(p_n-r+1)m})^T$,
$r=2$, $\ldots$, $p_n+1$. It is easy to see that
\[\hat{\md}_{r}+\hat{\md}_{r+1}=\left(\mX^T_{(r+2)}\mX_{(r+2)}\right)^{-1}\mX^T_{(r+2)}\my^{(r+2)}-
\left(\mX^T_{(r)}\mX_{(r)}\right)^{-1}\mX^T_{(r)}\my^{(r)}.\] It is
obvious that under $H_0$, for any $ \ell \in \{1,\ldots,p_n\}$ and
any $ i \in \{n-p_nm+1,\ldots,n\}$,
\[\bomega_\ell(i)=\mzero ~~\text{and} ~~ \md_{\ell}=\mzero.\]
We have the following theorem.

\emph{Theorem 1.} Assume that $m\to\infty$ as $n\to\infty$. If $H_0$
holds, under the assumptions C1-C2, it follows that
\[\sqrt{m}\hat{\md}_i \rightarrow_d
N\left(\mzero,2\sigma^2 W^{-1}\right),\quad\mbox{
$i=1,\ldots,p_n$.}\]

We now assume that $H_1$ holds. In view of (\ref{hh1}), it follows
that $\md_{r_k}+\md_{r_k+1}=\mdelta_{k}$. By the definition of
$\{\bomega_{\ell}(i)\}$, we have \bea
&&\sum_{\ell=1}^{p_n}\bomega_{\ell}(i)I(n-(p_n-\ell+1)m<i\leqslant
n-(p_n-\ell)m)\nonumber\\
&=&\left\{\begin{array}{ll} -\mdelta_k,&\hspace{.2in}\mbox{if
$\exists\:
r_k$ such that $n-(p_n-r_k+1)m<a_{k,n}<n-(p_n-r_k)m$},\\
&\label{omega}\\ \mzero,&\hspace{.2in}\mbox{otherwise.}
\end{array}\right.
\eea It can also be verified that \bea
&&\sum_{\ell=1}^{p_n}\md_{\ell}I(n-(p_n-\ell+1)m<i\leqslant
n)\nonumber\\&=&\left\{\begin{array}{ll}\displaystyle\sum_{\ell=1}^{r_k-1}\md_{\ell},
&\hspace{.2in}\mbox{if $n-(p_n-r_k+2)m<i\leqslant
n-(p_n-r_k+1)m$},\\
&\label{sumd}\\
\displaystyle\sum_{\ell=1}^{r_k+1}\md_{\ell}, &\hspace{.2in}\mbox{if
$n-(p_n-r_k)m<i\leqslant n-(p_n-r_k-1)m$}.
\end{array}\right.
\eea Thus, we have the following theorem:

\emph{Theorem 2.} If Assumptions C1-C2 hold, then under $H_1$,
\[\sqrt{m}\left(\hat{\md}_{r_k}+\hat{\md}_{r_k+1}-\mdelta_{k}\right)\rightarrow_d
N\left(\mzero,2\sigma^2 W^{-1}\right),\quad k=1,\ldots, K_0.\]

The proofs of Theorems 1-2 follow from the least squares theory. The
details are omitted.

\vspace{.2in} \noindent{\textsf{2.2. Estimate $\{\md_{i}\}$ by recent advances in consistent variable
selection methods}}

\vspace{.2in} \noindent{\emph{2.2.1. Estimate $\{\md_{i}\}$ by the adaptive LASSO}}

\noindent The adaptive LASSO, extending the LASSO in Tibshirani (1996), was proposed in Zhou (2006)
and possesses oracle properties for fixed number of regression coefficients.

In light of Zhou (2006), the adaptive LASSO type estimator of
$\mtheta_n$ for the model (\ref{reg}) is defined by
\begin{eqnarray}
\breve{\mtheta}_n=\arg\min_{\mtheta_n}\left\{\left|\left|\my-\mX_n\mtheta_n\right|\right|^2+
\lambda_n\sum_{r=1}^{p_n}\frac{1}{|{\tilde{\md}}_{r}|^\nu}
\left|\md_{r}\right|\right\},\label{gl}
\end{eqnarray}
where $\nu>0$, $\lambda_n$ is a thresholding parameter and ${\tilde{\md}}_{r}$
$\{r=1,\cdots,p_n\}$ are initial estimators satisfying
certain conditions.

\emph{Remark 3.} The adaptive LASSO estimate of $\mtheta_n$ may also
be defined by
\begin{eqnarray}
\check{\mtheta}_n=\arg\min_{\mtheta_n}\left|\left|\my-\mX_n\mtheta_n\right|\right|^2+
\lambda_n\sum_{r=1}^{p_n}\sum_{i=1}^{q}\frac{1}{|{\tilde{d}}_{ri}|^\nu}\left|d_{ri}\right|+
\gamma_n\sum_{i=1}^{q}\frac{1}{|{\tilde{\beta}}_{0i}|^\nu}\left|\beta_{0i}\right|,\label{al}
\end{eqnarray}
where $\mu>0$, $\lambda_n$ and $\gamma_n$ are thresholding
parameters satisfying certain conditions. The difference between
(\ref{gl}) and (\ref{al}) is that the variable selection in addition
to the multiple change point detection is also considered in
(\ref{al}). Due to the similarity in the techniques for finding the
asymptotic behavior of both $\breve{\mtheta}_n$ and
$\check{\mtheta}_n$, we only consider $\breve{\mtheta}_n$ in this
paper for simple presentation.

Since the dimension of $\mtheta_n$ increases with $n$ in
(\ref{reg}), the asymptotic results in Zhou (2006) are not
applicable here. In the following we will investigate the limiting
behavior of those $\md_i$s associated with change points under the
condition that $K_0\geqslant 1$, i.e., there exists at least one
change point in the model (\ref{cp}). As stated before, the
subscript $n$ may be suppressed for convenience if there is no
confusion.

Before we proceed, we define some notations as follows: Let
$\mB=\{\kappa_1,\kappa_2,\ldots,\kappa_\iota\}\subset
\{2,\ldots,p_n+1\}$ such that $\kappa_1<\ldots<\kappa_\iota$. Denote
$\mtheta_\mB=(\md^T_{\kappa_1},\cdots,\md^T_{\kappa_\iota})^T$,
$\mXB=(X_n^{(\kappa_1)}, \ldots, X_n^{(\kappa_\iota)})$, where
$\{X_n^{(i)}\}$ are given in (\ref{xx}).

Recall that for each $\mdelta_k$ in (\ref{cp}), there exists $r_k$
such that $\md_{r_k}=\mdelta_k$, or equivalently there exists a
change point within $\{n-(p_n-r_k+1)m,\ldots,n-(p_n-r_k)m-1\}$ for
$k=1,\ldots,K_0$.  Define
\begin{eqnarray}
&\ma_c=\{i:\ \md_{i-1}=\mzero, \ \md_i\neq\mzero, \
\md_{i+1}=\mzero\},&\ \ \ma_1=\{i:\ \md_{i-1}\neq\mzero, \
\md_i=\mzero ,
\ \md_{i+1}=\mzero\},\nonumber\\
&\ma_2=\{i:\ \md_{i-1}=\mzero, \ \md_i=\mzero, \
\md_{i+1}\neq\mzero\},&\ \ \ma_3=\{i:\ \md_{i-1}=\mzero, \
\md_i=\mzero,
 \ \md_{i+1}=\mzero\}.\nonumber
\end{eqnarray}
It is easy to see that for large $n$,
$\bar{\ma}_c=\ma_1\cup\ma_2\cup\ma_3.$

In view of Zhou (2006) and Huang, Ma and Zhang (2008), we need to
make some assumption on the initial estimators $\{\tilde{\md}_i\}$
used in (\ref{gl}) for investigating the asymptotic properties of
$\breve{\mtheta}_n$. By the remark 1 of Zhou (2006), one might
assume that for any $i$, there is a sequence of $\{a_n\}$ such that
$a_n\rightarrow\infty$ and $a_n(\tilde{\md}_i-\md_i)=O_p(1)$. But
$p_n$ is fixed in Zhou (2006).  Huang, Ma and Zhang (2008) allows
$p_n\to\infty$ as $n\to \infty$. Thus a stronger assumption like
that $r_n\max_i |\tilde{\md}_i-\md_i|=O_p(1)$ as
$r_n\rightarrow\infty$ (see (A2) of Huang, Ma and Zhang 2008) might
be made. However such assumptions may not be enough for the multiple
change point detection problem. A careful study shows that we need
put some lower bound on $|\tilde{\md}_i|$ for $i\in\ma_c$ such that
they are not close to $0$. Hence we make the following assumption on
$\{{\tilde{\md}}_{r}\}$:

\noindent{\bf Assumption C3.}\quad There exists a constant $a>0$
such that for large $n$,
\begin{eqnarray}
|\tilde{\md}_i|\begin{cases} \geqslant a>0, &\text{for $i\in\ma_c$,}\\
=O_p\left(1/\sqrt{m}\;\right), &\text{for $i\notin\ma_c$.}\\
\end{cases}\nonumber
\end{eqnarray}

To obtain $\{{\tilde{\md}}_{r}\}$ in practice, we can estimate the
set $\ma_c$ first, which, for example, may be estimated by the lease
squares based multiple change point detection algorithm given in
Subsection 3.1. After we obtain the estimate $\hat{\ma}_c$ of
${\ma}_c$, we can set $\tilde{\md}_i=c$ for $i\in\hat{\ma}_c$, and
${\bf{1}}_q/\sqrt{m}$ otherwise.

To study the asymptotic behavior of $\breve{\mtheta}$, the following
three Lemmas are necessary.

\emph{Lemma 1.}  Under Assumption C1, there exists positive definite
matrix $\mathcal{W}_{\ma_c}$ (defined in (\ref{wc}) in the appendix)
such that $X_{\ma_c}^TX_{\ma_c}/n\rightarrow \mathcal{W}_{\ma_c}$.

\emph{Remark 4.} One can not replace $X_{\ma_c}^TX_{\ma_c}$ by
$\tilde{X}_n^T\tilde{X}_n$ above since the minimum eigenvalue may
converge to 0 in consideration of the fact that $p_n\to\infty$ (see
Condition (b) in Zou (2006) and (2.13) in Zhang and Huang (2008)).
Thus if they allow $p_n\to\infty$, their conditions no longer hold
and may be strengthened as Assumption C1.

\emph{Lemma 2.}   Under Assumption C1, for large $n$  elements of
$\tilde{X}^T_{n}\mx_\omega/m$ are uniformly bounded.

\emph{Lemma 3.}   Under Assumptions C1-C2, for large $n$ elements of
$\tilde{X}_{n}^T\me_n/\sqrt{n}$ is uniformly bounded in probability.

If there exists at least one change point, i.e., $K_0\geqslant1$,
the limiting behavior of the adaptive LASSO estimator
$\breve{\mtheta}_n$ is given in the following theorem.

\emph{Theorem 3.} Assume that $\lambda_n/\sqrt{n}\rightarrow 0$,
$m/\sqrt{n}\rightarrow0$  and
$\lambda_n(n/p_n)^{\nu/2}/\sqrt{n}\rightarrow\infty$ for $\nu>0$ as
$n\to\infty$. If Assumptions C1-C3 hold, then
\[\sqrt{n}(\breve{\mtheta}_{\ma_{c}}-(\mtheta_n)_{\ma_c})\rightarrow_d
N(\mzero,\sigma^2\mathcal{W}_{\ma_c}^{-1}). \]

\emph{Remark 5.} \quad If we replace the weight $1/|x|^{\nu}$ by
$\exp(-1/|x|)$ in (\ref{gl}), the condition
$\lambda_n(n/p_n)^{\nu/2}/\sqrt{n}\rightarrow\infty$ can be relaxed
to the weaker condition:
$\lambda_n\exp\left(\sqrt{n/p_n}\;\right)/\sqrt{n}\to\infty$.
Although it may result in an absorbing state in $x=0$ (see Fan and
Lv (2008)), it has not occurred in simulations.

\emph{Remark 6.} \quad By (\ref{gl}), $\breve{\mtheta}$ is a unique
solution of a convex optimization problem and hence the
Karush-Kunh-Tucker condition holds. For any vector $\mb = (b_1,
\ldots, b_p)^T$, denote its sign vector by $\mbox{sgn}(\mb) =
(\mbox{sgn}(b_1),\ldots, \mbox{sgn}(b_p))^T$, with the convention
$\mbox{sgn}(0) = 0$. As in Zhao and Yu (2006), we say that
$\breve{\mtheta}_n=_s\mtheta$ if and only if
$\mbox{sgn}(\breve{\mtheta}_n)=\mbox{sgn}(\mtheta)$. If the
condition $p_n/n^{\nu/(2+\nu)}=o(1)$ is further assumed to hold, by
Lemma 1-3 and Theorem 3,  it can be shown that
\[P(\breve{\mtheta}_n=_s\mtheta)\to1,\quad\mbox{as $n\to\infty$.}\]
The proof is similar to the proof of Theorem 1 in Huang, Ma and
Zhang (2008) and hence omitted.

\vspace{.2in} \noindent{\emph{2.2.2. Estimate $\{\md_{i}\}$ by the SCAD or MCP}}

\noindent SCAD (Fan and Li (2001)) and MCP (Zhang (2010)) are two popular recent consistent variable
selection methods. They can also be employed to solve the multiple change point detection problem.

Consider the following estimator of $\mtheta_n$:
\begin{eqnarray}
\hat{\mtheta}^{{}_P}=\arg\min_{\mtheta}\left\{\left|\left|\my-\mX_n\mtheta\right|\right|^2+
n\sum_{r=1}^{p_n}p_{\lambda,\gamma}( \left|\md_r\right|)
\right\},\nonumber
\end{eqnarray}
where $p_{\lambda,\gamma}$ is the penalty function with tuning
parameters $\lambda>0$ and $\gamma>0$. If
\begin{eqnarray} p_{\lambda,\gamma}(x)=\left\{\begin{array}
{ll}\lambda x,& \text{if}~ x\leqslant
\lambda,\\
\displaystyle\frac{\gamma\lambda x-0.5(x^2+\lambda^2)}{\gamma-1} , &
\text{if}~ \lambda<x\leqslant \gamma\lambda,\\
\displaystyle\frac{\lambda^2(\gamma+1)}{2} , & \text{if}~ x>
\gamma\lambda,
\end{array}\right.\label{scad1}
\end{eqnarray}
the SCAD penalty function proposed by Fan and Li (2001),
${\hat{\mtheta}}^{{}_P}$ is the SCAD type estimator of $\mtheta_n$.
Denote it by $\hat{\mtheta}^{\SCAD}$. Instead, let
\begin{eqnarray} p_{\lambda,\gamma}(x)=\left\{\begin{array}
{ll}\lambda x-\frac{x^2}{2\gamma},& \text{if}~ x\leqslant
\gamma \lambda,\\
\frac{1}{2} \gamma\lambda^2, & \text{if}~ x>\gamma\lambda,
\end{array}\right.\label{mcp1}
\end{eqnarray}
the MCP penalty function proposed by Zhang (2010),
$\hat{\mtheta}^{{}_P}$ becomes the MCP type estimator of
$\mtheta_n$. Denote it by $\hat{\mtheta}^{\MCP}$.

Under certain conditions, the asymptotic properties of both
$\hat{\mtheta}^{\SCAD}$ and $\hat{\mtheta}^{\MCP}$ are similar to
the asymptotic properties of $\breve{\mtheta}$. Since the emphasis
of this paper is on the algorithms for detecting multiple change
points, their asymptotic properties will not be discussed here.

\vspace{.2in} \noindent{\textbf{3. Multiple change points detection algorithms}}

\noindent For a given $p_n$ or $m$, we divide the data sequence into $p_n+1$ segments such that the
first segment has the length between $m$ and $c_0m$ with $c_0\geqslant1$ and the rest $p_n$ segments
are all of length $m$, and we have the model (\ref{newcp}). Define
\be\hat{\sigma}_n^2=\sum_{\ell=1}^{n-p_nm}(y_\ell-\mx_\ell^T\hat{\mbeta})^2/(n-p_nm-q)\label{sigma}\ee
with $\hat{\mbeta}=(\mX^T_{(1)}\mX_{(1)})^{-1}\mX^T_{(1)}\my^{(1)}$. Given a significance level
$\alpha$, five multiple change point detection algorithms are proposed in this section.

\vspace{.2in} \noindent{\textsf{3.1 Least squares based multiple change point detection algorithm}}

\noindent In light of Theorems 1-2, the least squares based multiple change point detection algorithm
is given as follows:

\no{\emph{L}east \emph{s}quares based \emph{m}ultiple \emph{c}hange
\emph{p}oints \emph{d}etection \emph{a}lgorithm (LSMCPDA):}

\noindent{\emph{Step 1.}} Set $i=1$, $j=1$ and $\hat{K}=0$.

\noindent{\emph{Step 2.}}  If $i\geqslant p_n-3$, go to Step 3.
Otherwise, we test the hypothesis $H_{0,i}:\;\md_{i}=\mzero$ by
checking if
\[
\hat{\md}{}^T_{i}\mX_{(i+1)}^T\mX_{(i+1)}\hat{\md}_{i}/(2q\hat{\sigma}_n^2)\geqslant
\chi^2_{\alpha,q},\] where $\hat{\md}_i$ is given in (\ref{ddd}). If
the test is significant, set $i=i+1$ and repeat Step 2, otherwise we
test the hypothesis $H_{0,(i+1,i+2)}:\;\md_{i+1}+\md_{i+2}=\mzero$
by checking if
\[\left(\hat{\md}_{i+1}+\hat{\md}_{i+2}\right)^T\mX_{(i+1)}^T
\mX_{(i+1)}\left(\hat{\md}_{i+1}+\hat{\md}_{i+2}\right)/(2q\hat{\sigma}_n^2)\geqslant
\chi^2_{\alpha,2q}.\] If the test is not significant, set $i=i+1$
and repeat Step 2,  otherwise, a change point estimate is
$n-p_nm+im$. Set $\hat{r}_j=n-p_nm+im$, $j=j+1$, $i=i+2$, and
$\hat{K}=\hat{K}+1$. Then repeat Step 2.

\noindent{\emph{Step 3.}} If $\hat{K}=0$, then go to the next step.
Otherwise, we use the CUSUM to improve the accuracy of the multiple
change point detection as follows: We search for the change points
within the ${\hat{K}}$ sets: $\big\{\{n-p_nm+(\hat{r}_j-1)m,\ldots,
n-p_nm+(\hat{r}_j+1) m\},\ j=1,\ldots,\hat{K}\big\}$ by the CUSUM.
An estimate of the change point within the $j$th set is given by
\[\hat{a}_{j,n}=\arg\max_\ell\left[\min_{\mbeta}\sum_{j=n-p_nm+(\hat{r}_j-1)m}^\ell(y_j-\mx_j^T\mbeta)^2
+\min_{\mbeta}\sum_{j=\ell+1}^{n-p_nm+(\hat{r}_j+1)
m}(y_j-\mx_j^T\mbeta)^2\right].\]

\noindent{\emph{Step 4.}} If $\hat{K}=0$, there is no change points.
Otherwise, there are ${\hat{K}}$ change points and they are
$\hat{a}_{1,n},\ldots, \hat{a}_{\hat{K},n}$.

If in the algorithm above, the chi-square tests in Step 2 are
replaced by the CUSUM tests (see Appendix A.1) and Step 3 is
replaced by Steps 3-5 of the SMCPDA with $\{\hat{r}^{\SCAD}_j\}$,
$\{\hat{\md}_j^{\SCAD}\}$, $\hat{K}^{\SCAD}$ and
$\{\hat{a}^{\SCAD}_{j,n}\}$ replaced by $\{\hat{r}_j\}$,
$\{\hat{\md}_j\}$, $\hat{K}$, and $\{\hat{a}_{j,n}\}$ respectively,
the new algorithm is named as CLSMCPDA, where ``$C$'' is the first
letter of ``\emph{C}USUM''.

\vspace{.2in} \noindent{\textsf{3.2 Adaptive LASSO based multiple change poins detection algorithm}

\noindent In light of Theorems 3, the adaptive LASSO based multiple change point detection algorithm
is given as follows:

\no{\emph{A}daptvie \emph{L}asso based \emph{m}ultiple \emph{c}hange
\emph{p}oints \emph{d}etection \emph{a}lgorithm (ALMCPDA):}

\noindent{\emph{Step 1.}} Set $i=1$, $j=1$ and $\breve{K}=0$.
Execute the algorithm LSMCPDA and obtain $\hat{K}$. If $\hat{K}>0$,
we also obtain $\hat{a}_{1,n},\ldots, \hat{a}_{\hat{K},n}$.

\vspace{.2in}\noindent{\emph{Step 2.}}  If $\hat{K}=0$, set
$\tilde{\md}_1=\cdots=\tilde{\md}_{p_n}={\bf{1}}_q/\sqrt{m}$,
otherwise, set
\begin{eqnarray}
\tilde{\md}_\ell=\begin{cases} c{\bf{1}}_q,& \ell\in\{r_k,\ r_km<\hat{a}_{k,n}-n+p_nm\leqslant (r_k+1)m\},\\
{\bf{1}}_q/\sqrt{m},& \mbox{elsewhere};
\end{cases}\nonumber
\end{eqnarray}
where $r_k$ is an integer such that
$r_km<\hat{a}_{k,n}-n+p_nm\leqslant (r_k+1)m$ and $c$ is a prechosen
constant. Select $\lambda>0$ and $\nu>0$. Find the adaptive LASSO
estimate $\breve{\mtheta}$ of $\mtheta$ via
\begin{eqnarray}
\breve{\mtheta}=\arg\min_{\mtheta}\left\{\left|\left|\my-\mX_n\mtheta\right|\right|^2+
\lambda\sum_{r=1}^{p_n}\frac{1}{|{\tilde{\md}}_{r}|^\nu}
\left|\md_{r}\right|\right\},\nonumber
\end{eqnarray}
and we obtain $\breve{\md}_\ell$ for $1\leqslant \ell\leqslant p_n$.

\vspace{.2in} \noindent{\emph{Step 3.}} We compute
$z_\ell=||\breve{\md}_\ell||_\infty$ for $1\leqslant \ell\leqslant
p_n$. If $z_1=z_2=\cdots=z_{p_n}=0$, go to Step 5. Otherwise, we
treat $\{z_\ell\}$ as random variables from the model
$\mz=\bmu+\mee$ with $\bmu=(\mu_1,\ldots,\mu_{p_n})^T$ and $\mee\sim
N({\bf{0}},I_{p_n})$. Use LASSO, SCAD or MCP among other recent
advances in variable selection to perform variable selection based
on $\{z_{\ell}\}$. We obtain the estimates $\{\tilde{\mu}_{\ell}\}$.
If $\tilde{\mu}_\ell$, $1\leqslant\ell\leqslant p_n$, are all zeros,
set $\breve{K}=0$ and go to Step 6. Otherwise, let $\mI$ be the
subset of $\{1,\ldots,p_n\}$ such that $\ell\in\mI$ if and only if
$\tilde{\mu}_{\ell}\not=0$. Write $\mI=\{s_1,\dots,s_{|\mI|}\}$ such
that $s_1<\ldots<s_{|\mI|}$.

\noindent{\emph{Step 4.}} If $i> |\mI|$, go to Step 5. Otherwise, we
test the hypothesis $H_{0,s_i}:\;\md_{s_i}=\mzero$ by checking if
\[
(p_n-s_i)\breve{\md}{}^T_{s_i}\mX_{(s_i+1)}^T\mX_{(s_i+1)}\breve{\md}_{s_i}/(q\hat{\sigma}_n^2)\geqslant
\chi^2_{\alpha,q},\] where $\hat{\sigma}_n^2$ is given in
(\ref{sigma}). If the test is not significant, set $i=i+1$ and
repeat Step 4. Otherwise, a change point estimate is
$n-p_nm+(s_i-1)m$. Set $\breve{r}_j=n-p_nm+(s_i-1)m$, $j=j+1$,
$i=i+2$, and $\breve{K}=\breve{K}+1$. Then repeat Step 4.

\noindent{\emph{Step 5.}} If $\breve{K}=0$, then go to the next
step. Otherwise, we use the CUSUM to improve the accuracy of the
multiple change point detection as follows: We search for the change
points within the ${\breve{K}}$ sets:
$\big\{\{n-p_nm+(\breve{r}_j-1)m,\ldots, n-p_nm+(\breve{r}_j+1)
m\},\ j=1,\ldots,\breve{K}\big\}$ by the CUSUM. An estimate of the
change point for the $j$th set is given by
\[\breve{a}_{j,n}=\arg\max_\ell\left[\min_{\mbeta}\sum_{j=n-p_nm+(\breve{r}_j-1)m}^\ell(y_j-\mx_j^T\mbeta)^2
+\min_{\mbeta}\sum_{j=\ell+1}^{n-p_nm+(\breve{r}_j+1)
m}(y_j-\mx_j^T\mbeta)^2\right].\]

\noindent{\emph{Step 6.}} If $\breve{K}=0$, there is no change
points. Otherwise, there are ${\breve{K}}$ change points and they
are $\breve{a}_{1,n},\ldots, \breve{a}_{\breve{K},n}$.

If the algorithm above, the chi-square test is replaced by the CUSUM
test in Step 4, the new algorithm is named as CALMCPDA, where
``$C$'' is also the first letter of ``\emph{C}USUM''. Denote all the
estimates based on CALMCPDA by adding a superscript ``C'' to the
corresponding estimates based on ALMCPDA. For example, the estimate
of $K_0$ based on CALMCPDA is denoted by $\breve{K}^C$.

\vspace{.2in} \noindent{\textsf{3.3 SCAD based multiple change points detection algorithm}

\noindent Similar to the ALMCPDA, the SCAD based multiple change point detection algorithm is given as
follows:

\no{\emph{S}CAD  based \emph{m}ultiple \emph{c}hange \emph{p}oints
\emph{d}etection \emph{a}lgorithm (SMCPDA):}

\noindent{\emph{Step 1.}} Set $i=1$, $j=1$ and $\hat{K}^{\SCAD}=0$.

\vspace{.2in}\noindent{\emph{Step 2.}}  Select $\lambda>0$ and
$\gamma>0$. Find the SCAD estimate
$\hat{\mtheta}^{\SCAD}=\left(\left(\hat{\mbeta}^{\SCAD}\right)^T\right.$,
$\left(\hat{\md}_1^{\SCAD}\right)^T$, $\ldots$,
$\left.\left(\hat{\md}_{p_n}^{\SCAD}\right)^T\right)^T$ of $\mtheta$
via
\begin{eqnarray}
\hat{\mtheta}^{\SCAD}=\arg\min_{\mtheta}\left\{\left|\left|\my-\mX_n\mtheta\right|\right|^2+
n\sum_{r=1}^{p_n}p_{\lambda,\gamma}( \left|\md_r\right|)
\right\},\nonumber
\end{eqnarray}
where $p_{\lambda,\gamma}$ is given in (\ref{scad1}) and we obtain
$\hat{\md}_\ell^{\SCAD}$ for $1\leqslant \ell\leqslant p_n$.

\vspace{.2in} \noindent{\emph{Step 3.}} It is same as Step 3 of
ALMCPDA with $z_\ell=||\breve{\md}_\ell||_\infty$ is replaced by
$z_\ell=\left\|\hat{\md}_\ell^{\SCAD}\right\|_\infty$ for
$1\leqslant \ell\leqslant p_n$ and $\breve{K}=0$ is replaced by
${\hat{K}}^{\SCAD}$.

\noindent{\emph{Step 4.}} If $i> |\mI|$, go to Step 5. Otherwise, we
test the hypothesis $H_{0,s_i}:\;\md_{s_i}=\mzero$ by CUSUM. If the
test is not significant, set $i=i+1$ and repeat Step 4.  Otherwise,
a change point estimate is $n-p_nm+(s_i-1)m$. Set
$\hat{r}^{\SCAD}_j=n-p_nm+(s_i-1)m$, $j=j+1$, $i=i+2$, and
$\hat{K}^{\SCAD}=\hat{K}^{\SCAD}+1$. Then repeat Step 4.

\noindent{\emph{Step 5.}} If $\hat{K}^{\SCAD}=0$, then go to the
next step. Otherwise, we use the CUSUM to improve the accuracy of
the multiple change point detection as follows: We search for the
change points within the ${\hat{K}^{\SCAD}}$ sets:
$\big\{\{n-p_nm+(\hat{r}^{\SCAD}_j-1)m,\ldots,
n-p_nm+(\hat{r}^{\SCAD}_j+1) m\},\ j=1,\ldots,\hat{K}^{\SCAD}\big\}$
by the CUSUM. An estimate of the change point for the $j$th set is
given by
\[\hat{a}^{\SCAD}_{j,n}=\arg\max_\ell\left[\min_{\mbeta}\sum_{j=n-p_nm+(\hat{r}^{\SCAD}_j-1)m}^\ell(y_j-\mx_j^T\mbeta)^2
+\min_{\mbeta}\sum_{j=\ell+1}^{n-p_nm+(\hat{r}^{\SCAD}_j+1)
m}(y_j-\mx_j^T\mbeta)^2\right].\]

\noindent{\emph{Step 6.}} If $\hat{K}^{\SCAD}=0$, there is no change
points. Otherwise, there are $\hat{K}^{\SCAD}$ change points and
they are $\hat{a}^{\SCAD}_{1,n},\ldots,
\hat{a}^{\SCAD}_{\hat{K}^{\SCAD},n}$.

\vspace{.2in} \noindent{\textsf{3.4 MCP based multiple change points detection algorithm}

\noindent The differences between the SMCPDA and the MCP based multiple change point detection
algorithm (MMCPDA) are as follows:

\begin{enumerate}
\item The superscript ``$scad$'' in the SMCPDA is replaced by the
superscript ``$mcp$'' in the MMCPDA.

\item The step 2 in the SMCPDA is modified to the following step 2 in the MMCPDA:

\noindent{\emph{Step 2.}}  Select $\lambda>0$ and $\gamma>0$. Find
the MCP estimate
$\hat{\mtheta}^{\MCP}=\left(\left(\hat{\mbeta}^{\MCP}\right)^T\right.$,
$\left(\hat{\md}_1^{\MCP}\right)^T$, $\ldots$,
$\left.\left(\hat{\md}_{p_n}^{\MCP}\right)^T\right)^T$ of $\mtheta$
via
\begin{eqnarray}
\hat{\mtheta}^{\MCP}=\arg\min_{\mtheta}\left\{\left|\left|\my-\mX_n\mtheta\right|\right|^2+
n\sum_{r=1}^{p_n}p_{\lambda,\gamma}( \left|\md_r\right|)
\right\},\nonumber
\end{eqnarray}
where $p_{\lambda,\gamma}$ is given in (\ref{mcp1}).

\end{enumerate}

\emph{Remark 7.} \quad  The use of CUSUM in these algorithms is for
improving the change point estimation accuracy. The amounts of
computing time required by these algorithms are all $O(n)+O(m)$,
where $O(m)$ corresponds to the time required for using CUSUM
method.  If a segmentation satisfies that $m=o(n)$,
$O(n)+O(m)=O(n)$, which is computationally more efficient than the
existing multiple change point detection methods in literature.}

\vspace{.2in} \noindent{\textbf{4. Simulation study}}

\noindent In this section, we present simulation studies of multiple change point analysis. Since the
time for finding the multiple change points in a large sample by the algorithms proposed in Section 3
is significantly reduced compared to the existing multiple change point detection methods in the
literature, such comparison studies are omitted in this section. We will only compare the number of
times of selecting the true number of change points and the accuracy of change point estimation by the
algorithms proposed in Section 3 based on 1000 simulation. A Dell server (two E5520 Xeon Processors,
two 2.26GHz 8M Caches, 16GB Memory) is used in the simulation.

It is noted that the LARS algorithm (Efron, Hastie, Johnstone, and
Tibshirani 2004) is used to compute $\breve{\mtheta}_n$ defined in
(\ref{gl}) with $\nu=1$ and an optimal $\lambda_n$ selected by the
BIC. For applying LARS, the added penalty on $\mbeta$ is set as
$1/|{\bf{1}}_q|$, which will not affect the multiple change-point
detection results as $\mbeta\not=\mzero$. The PLUS algorithm (Zhang,
2010) with the added penalty $np_{\lambda,\gamma}(
\left|\mbeta\right|)$ on $\mbeta$ is used to compute
$\hat{\mtheta}_n^\SCAD$ defined in (\ref{scad1}) or
$\hat{\mtheta}_n^\MCP$ defined in (\ref{mcp1}), which also do not
affect the multiple change point detection results as
$\mbeta\not=\mzero$. Let $\hat{\sigma}_n^2$ be given in
(\ref{sigma}). We use $\lambda=\hat{\sigma}_n\sqrt{2\log p_n/n}$ in
the PLUS algorithm as suggested in Zhang (2010). In all of our
numerical examples, we set $\gamma=3.7$ for SCAD by following the
recommendation of Fan and Li (2001), but set $\gamma=2.4$ for MCP
based on some preliminary simulation studies. It is noted that in
the step 3 of the algorithms ALMCPDA, CALMCPDA, SMCPDA, and MMCPDA,
we use SCAD to perform variable selection for model $\mz=\bmu+\mee$
by applying the PLUS algorithm with $\lambda=0.02$. To use such
small $\lambda$ is for avoiding the possibility of overestimation of
the number of multiple change points.

Throughout this section, $\alpha=0.05$.

\vspace{.2in} \noindent{\textsf{4.1. The case that there is no change point in the data sequence of
size 5000}}

\noindent In this subsection, we consider the case that there is no change point in the data sequence.
We will examine the performance of the proposed algorithms to see if they do claim that there is no
change point.

Consider the following linear model
\[
y_i=\mx_i^T\mbeta_0+\varepsilon_i,\quad i=1,\ldots, n,
\]
where $\mbeta_0$ is a $q\times1$ parameter vector. Set $n=5000$,
$q=3$, $\mbeta_0=(1, 1.4, 0.7)^T$, and $x_{i,1,n}=1$ for
$i=1,\ldots,5000$. Generate $\varepsilon_i$, $i=1,\ldots,5000$, such
that they are i.i.d. $N(0,1)$ distributed, and generate two
sequences $x_{i,2,n}$, $1,\ldots,n$, and $x_{i,3,n}$,
$1,\ldots,5000$, such that they are i.i.d. $N(1,2)$ distributed. For
demonstration, a sample scatter plot of simulated data is given in
Figure 1.

\begin{figure}[!ht]
   \centering
   \includegraphics[width=1\textwidth]{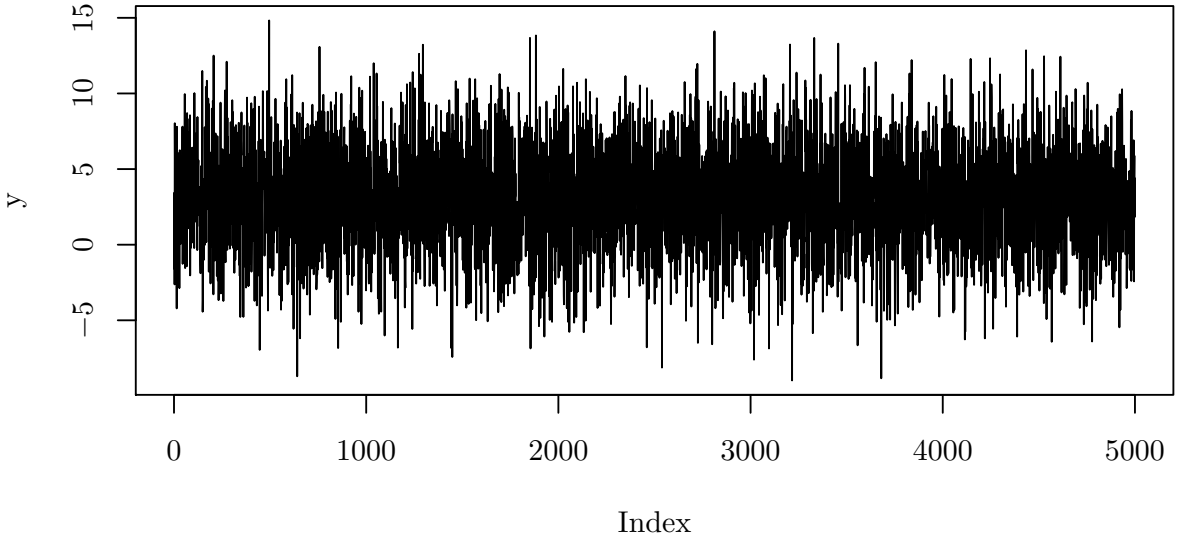}
   \caption{There is no change point in the data sequence.}
            \label{fig1}
 \end{figure}

We compare the following five algorithms: LSMCPDA, both ALMCPDA and
CALMCPDA with $c=1$, SMCPDA and MMCPDA. Recall that all the tests
used in the algorithms CALMCPDA, SMCPDA, and MMCPDA are based on
CUSUM. The number of correct detection and average computation time
in second based on 1000 simulations are given Table 1.

\vspace{.2in}

\renewcommand{\tabcolsep}{1mm}
\renewcommand{\arraystretch}{1.2}

\begin{table}[h] \caption{The entries are the numbers of correct change point
detection by the five algorithms LSMCPDA,  ALMCPDA,  CALMCPDA, SMCPDA and MMCPDA and the corresponding
average computation time based on 1000 simulations.}
\begin{center}
\begin{tabular} {|c|c|c|c|c|c|c|}
\hline
&LSMCPDA&ALMCPDA&CALMCPDA&SMCPDA&MMCPDA \\
\hline No. of Correct Detection &999&996& 1000 &1000&1000\\
\hline Average Computation Time &1.42&3.84&6.78&1.93&1.98\\
\hline
\end{tabular}
\end{center}

\end{table}

\noindent From Table 1, it can be seen that all algorithms perform
very well. The average detection time required by CALMCPDA for a
sample of size 5000 is more than other proposed algorithms but only
6.78 seconds.

\vspace{.2in} \noindent{\textsf{4.2. The case that there are nine change points in the data sequence
of size 5000}}

\noindent In this subsection, we consider a case that there are nine change points in the data
sequence of size 5000. We will examine the performance of the proposed algorithms via the rate for
correctly estimating the number of change points and the accuracy of change point estimation. The
average computation time for multiple change point detection is also given for each algorithm.

Consider the model (\ref{cp}), i.e., \beaa y_{i,n}&=&\sum_{j=1}^q
x_{i,j,n}\beta_{j,0}+\sum_{\ell=1}^{K_0}\sum_{j=1}^q
x_{i,j,n}\delta_{j,0}^{(\ell)}I(a^{(0)}_{\ell,n}<
i\leqslant n)+\varepsilon_{i,n}\nonumber\\
&=&\mx_{i,n}^T\left[\mbeta_0+\sum_{\ell=1}^{K_0}\mdelta_{\ell,0}
I(a^{(0)}_{\ell,n}< i\leqslant n)\right]+\varepsilon_{i,n},\quad
i=1,\ldots, n. \eeaa As in Subsection 4.1, set $n=5000$, $q=3$,
$\mbeta_0=(1, 1.4, 0.7)^T$, choose $p_n=\lfloor n/50\rfloor$ and
$m=\lfloor n/(p_n+1)\rfloor$, and generate $\{x_{i,j,n}\}$ and
$\{\varepsilon_i\}$ in the same way as in Subsection 4.1. Set
$K_0=9$, $\mdelta_1=\mdelta_3=\mdelta_5=\mdelta_7=\mdelta_9=(0.5,
-0.7, 0.4)^T$, and
$\mdelta_2=\mdelta_4=\mdelta_6=\mdelta_8=-\mdelta_1$. Consider the
following two change point location settings:

\begin{itemize}
\item[] CPL1. $a_{i}=500\times i$, for $i=1,\ldots,9$;

\item[] CPL2. $a_{1}=503$, $a_{2}= 923$, $a_{3}= 1471$, $a_{4}= 2077$, $a_{5}= 2334$, $a_{6}= 2890$,
$a_{7}=3410$, $a_{8}= 3909$, and $a_{9}= 4546$.

\end{itemize}
For demonstration, two scatter plots of simulated data for the
settings CPL1 and CPL2 are given respectively in Figures 2-3. One
can hardly find any change points from these two figures.

 \begin{figure}[!ht]
   \centering
   \includegraphics[width=1\textwidth]{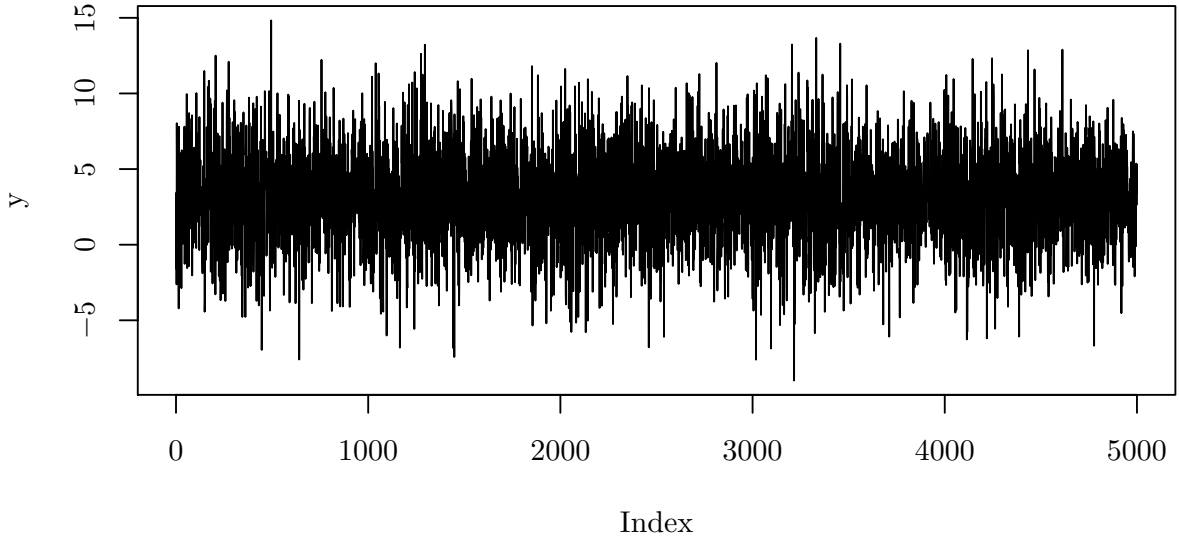}
   \caption{The scatter plot of simulated data for Setting CPL1.}
            \label{fig2}
 \end{figure}

 \begin{figure}[!ht]
   \centering
   \includegraphics[width=1\textwidth]{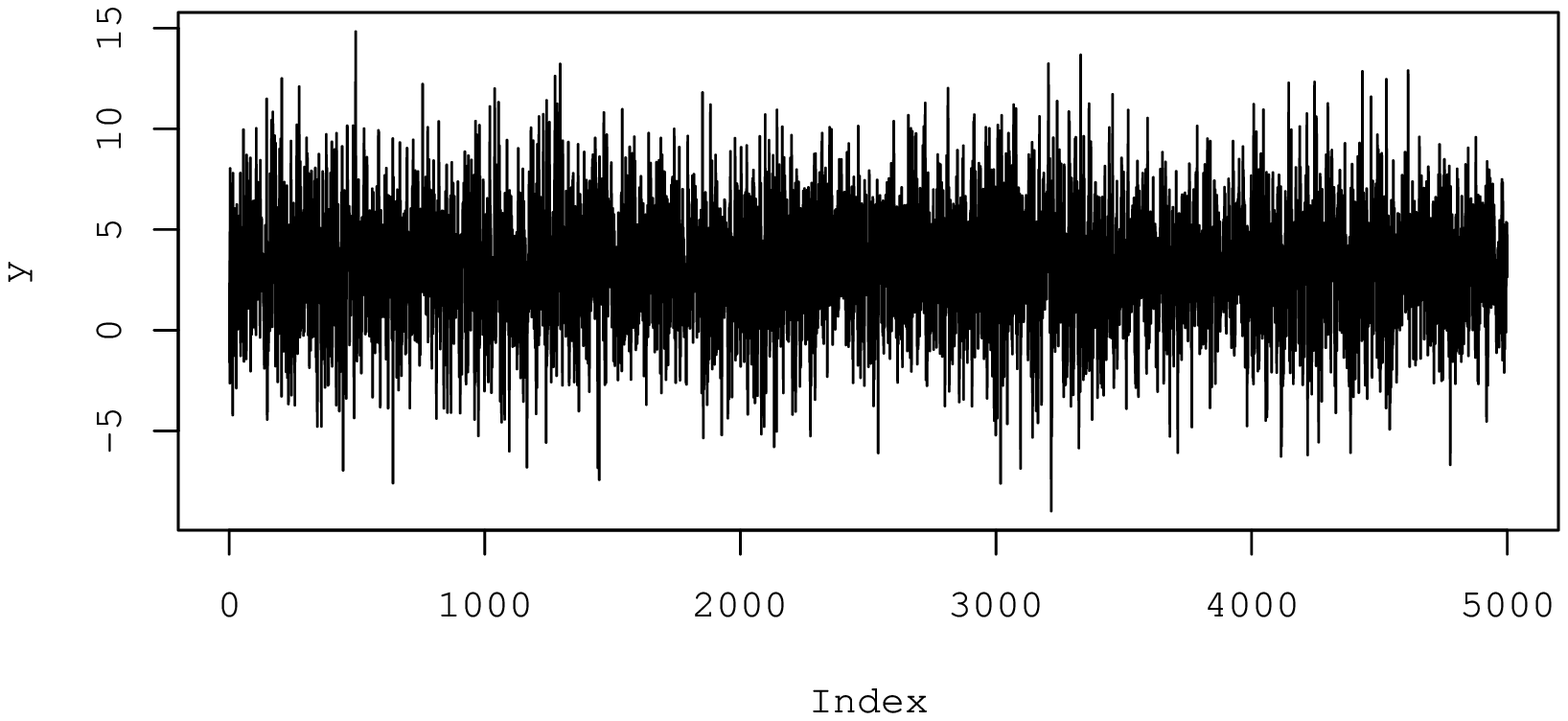}
   \caption{The scatter plot of simulated data for Setting CPL2.}
            \label{fig3}
 \end{figure}

We compare the following five algorithms: LSMCPDA, ALMCPDA,
CALMCPDA, SMCPDA and MMCPDA. Let $\tilde{a}_{i}$ stand for
$\hat{a}_{i}$, $\breve{a}_{i}$, $\breve{a}^{C}_{i}$,
$\hat{a}^{\SCAD}_{i}$ or $\hat{a}^{\MCP}_{i}$ for $i=1,\ldots,9$. We
check the accuracy of multiple change point estimation based on each
algorithm by examining the distance between $\tilde{a}_{i}$ and
${a}_{i}$ for $i=1,\ldots,9$. We only consider such distance to be
equal to 0 or less than or equal to 5 or 10. The simulation results
for the two change point location settings CPL1 and CPL2 are
presented in Tables 2-3.

\renewcommand{\tabcolsep}{1mm}
\renewcommand{\arraystretch}{1.2}
\begin{table}[ht] \caption{The entries are the numbers of $\tilde{a}_{i}$ such that
$|\tilde{a}_{i}-a_{i,n}|\leqslant0, 5, 10$ for $i=1,\ldots,9$, the number of correctly estimating the
number of change points and the corresponding average computation time by each of the five algorithms
LSMCPDA, ALMCPDA,  CALMCPDA, SMCPDA and MMCPDA based on 1000 simulations for the change point location
setting CPL1.}

\begin{center}

\begin{tabular} {|c|c|c|c|c|c|}\hline
&LSMCPDA&ALMCPDA&CALMCPDA&SMCPDA&MMCPDA  \\ \hline
{$|\tilde{a}_{1,n}-a_{1}|=0$}&208&215&215&212&212\\
{$|\tilde{a}_{1}-a_{1}|\leqslant5$}&958&973&973&973&974 \\
{$|\tilde{a}_{1}-a_{1}|\leqslant10$}&990&993&993&992&992 \\
\hline
$|\tilde{a}_{2}-a_{2}|=0$&489&532&532&520&525\\
$|\tilde{a}_{2}-a_{2}|\leqslant5$&924&939&939&918&922 \\
$|\tilde{a}_{2}-a_{2}|\leqslant10$&979&982&982&960&964 \\
\hline
$|\tilde{a}_{3}-a_{3}|=0$&263&262&262&253&253\\
$|\tilde{a}_{3}-a_{3}|\leqslant5$&806&807&807&773&792 \\
$|\tilde{a}_{3}-a_{3}|\leqslant10$&972&977&977&932&952 \\
\hline
$|\tilde{a}_{4}-a_{4}|=0$&162&174&174&157&174\\
$|\tilde{a}_{4}-a_{4}|\leqslant5$&810&806&806&773&786 \\
$|\tilde{a}_{4}-a_{4}|\leqslant10$&961&959&959&921&939 \\
\hline
$|\tilde{a}_{5}-a_{5}|=0$&716&726&726&694&703\\
$|\tilde{a}_{5}-a_{5}|\leqslant5$&961&975&975&931&947 \\
$|\tilde{a}_{5}-a_{5}|\leqslant10$&986&998&998&953&970 \\
\hline
$|\tilde{a}_{6}-a_{6}|=0$&210&223&223&215&218\\
$|\tilde{a}_{6}-a_{6}|\leqslant5$&980&985&985&941&956 \\
$|\tilde{a}_{6}-a_{6}|\leqslant10$&993&1000&1000&955&971 \\
\hline
$|\tilde{a}_{7}-a_{7}|=0$&201&219&219&195&204\\
$|\tilde{a}_{7}-a_{7}|\leqslant5$&824&876&876&814&844 \\
$|\tilde{a}_{7}-a_{7}|\leqslant10$&928&973&973&904&937 \\
\hline
$|\tilde{a}_{8}-a_{8}|=0$&455&511&511&460&474\\
$|\tilde{a}_{8}-a_{8}|\leqslant5$&893&978&978&897&927 \\
$|\tilde{a}_{8}-a_{8}|\leqslant10$&907&991&991&911&942 \\
\hline
$|\tilde{a}_{9}-a_{9}|=0$&240&277&277&276&279\\
$|\tilde{a}_{9}-a_{9}|\leqslant5$&786&935&936&922&918 \\
$|\tilde{a}_{9}-a_{9}|\leqslant10$&822&980&981&966&961 \\
\hline
{No. of Correct Detection} &818&950&987&898&920  \\
\hline
{Average Computation Time}&2.23&5.61&8.20&2.88&2.98 \\

\hline
\end{tabular}
\end{center}
\end{table}
\renewcommand{\tabcolsep}{1mm}
\renewcommand{\arraystretch}{1.2}
\begin{table}[h] \caption{The entries are the numbers of $\tilde{a}_{i}$ such that
$|\tilde{a}_{i}-a_{i,n}|\leqslant0, 5, 10$ for $i=1,\ldots,9$, the number of correctly estimating the
number of change points and the corresponding average computation time by each of the five algorithms
LSMCPDA, ALMCPDA,  CALMCPDA, SMCPDA and MMCPDA based on 1000 simulations for the change point location
setting CPL2.}
\begin{center}

\begin{tabular} {|c|c|c|c|c|c|}\hline
&LSMCPDA&ALMCPDA&CALMCPDA&SMCPDA&MMCPDA \\ \hline
$|\tilde{a}_{1}-a_{1}|=0$ &362&378&378&377&381 \\
$|\tilde{a}_{1}-a_{1}|\leqslant5$&955&961&961&955&960 \\
$|\tilde{a}_{1}-a_{1}|\leqslant10$&986&991&991&985&991 \\
\hline
$|\tilde{a}_{2}-a_{2}|=0$&270&276&275&271&274\\
$|\tilde{a}_{2}-a_{2}|\leqslant5$&858&872&869&861&865 \\
$|\tilde{a}_{2}-a_{2}|\leqslant10$&975&991&988&976&981 \\
\hline
$|\tilde{a}_{3}-a_{3}|=0$&426&522&522&522&523 \\
$|\tilde{a}_{3}-a_{3}|\leqslant5$&767&952&952&957&958 \\
$|\tilde{a}_{3}-a_{3}|\leqslant10$&811&982&982&987&988 \\
\hline
$|\tilde{a}_{4}-a_{4}|=0$&195&194&194&115&150\\
$|\tilde{a}_{4}-a_{4}|\leqslant5$&892&911&911&525&714 \\
$|\tilde{a}_{4}-a_{4}|\leqslant10$&955&970&970&562&766 \\
\hline
$|\tilde{a}_{5}-a_{5}|=0$&272&295&294&169&249\\
$|\tilde{a}_{5}-a_{5}|\leqslant5$&910&980&978&578&834 \\
$|\tilde{a}_{5}-a_{5}|\leqslant10$&921&997&995&582&845 \\
\hline
$|\tilde{a}_{6}-a_{6}|=0$&793&795&795&783&779\\
$|\tilde{a}_{6}-a_{6}|\leqslant5$&967&971&968&954&946 \\
$|\tilde{a}_{6}-a_{6}|\leqslant10$&987&993&988&972&964 \\
\hline
$|\tilde{a}_{7}-a_{7}|=0$&293&317&315&309&309\\
$|\tilde{a}_{7}-a_{7}|\leqslant5$&922&941&939&932&931 \\
$|\tilde{a}_{7}-a_{7}|\leqslant10$&973&991&989&984&986 \\
\hline
$|\tilde{a}_{8}-a_{8}|=0$&197&210&196&211&206\\
$|\tilde{a}_{8}-a_{8}|\leqslant5$&836&899&899&904&910 \\
$|\tilde{a}_{8}-a_{8}|\leqslant10$&891&968&968&969&975 \\
\hline
$|\tilde{a}_{9}-a_{9}|=0$&305&298&298&304&304\\
$|\tilde{a}_{9}-a_{9}|\leqslant5$&927&924&924&934&932 \\
$|\tilde{a}_{9}-a_{9}|\leqslant10$&974&977&977&982&982 \\
\hline
{No. of Correct Detection} &895&947&964&572&759  \\
\hline
{Average Computation Time}&2.29&5.97&8.65&3.00&2.98 \\

\hline
\end{tabular}
\end{center}
\end{table}

From both tables, it can be seen that all algorithms perform well in
terms of accuracy of multiple change point estimation and the rate
for correctly estimating the number of change points. The ALMCPDA
and CALMCPDA are compatible and in generally outperform others. The
average detection time required by CALMCPDA for a sample of size
5000 is more than all other algorithms, which is 8.20 seconds for
CPL1 and 8.65 seconds for CPL2. In contrast, the average detection
time required by ALMCPDA is only 5.61 seconds for CPL1 and 5.97
seconds for CPL2.

\noindent{\textsf{4.3. Practical recommendation of $p_n$}}

\noindent It is clear that the choice of $p_n$ will affect the performance of the proposed algorithms.
Too large $p_n$ may tend to underestimate the true number of multiple change points and increase
biases in change point estimation while may cut down the computation time. Hence a care must be taken
in choosing a proper $p_n$, and we propose the following algorithm:

\noindent{\emph{Step 1.}} We choose an initial set $\mB$ containing
probable values of $p_n$.

\noindent{\emph{Step 2.}}  For each $p_n$ in the set $\mB$, we
obtain an estimate of $\mtheta_n$ in (\ref{al}) by using an
algorithm, say ALMCPDA. We can then calculate the residual sum of
squares, denoted by $RSS(p_n)$.

\noindent{\emph{Step 3.}} The optimal $p_n$ is chosen as
$\arg\min_{p_n\in\mB} RSS(p_n)$.

\vspace{.2in} \noindent{\textbf{5. Empirical applications}}

\noindent In this section, we consider empirical applications of the multiple change point detection
methods proposed in this paper by analyzing the U.S. Ex-Post Real Interest Rate (Garcia and Perron,
1996) and Gross domestic product in U.S.A (Maddala, 1977).

\vspace{.2in} \noindent{\textsf{5.1. The U.S. Ex-Post Real Interest Rate}}

\noindent Garcia and Perron (1996) considered the time series behavior of the U.S. Ex-Post real
interest rate (constructed from the three-month treasury bill rate deflated by the CPI inflation rate
taken from the Citibase data base). The data are quarterly series from January, 1961 to March, 1986,
which is plotted in Figure 4. We are interested in finding out if there are change points in the mean
of the series. Thus we apply the proposed algorithms to the mean shift model. It is noted that by
Remark 2, the algorithms are applicable even if there exists potential serial correlation.

\begin{figure}[!ht]
  \centering
    \includegraphics[width=1\textwidth,height=0.38\textheight]{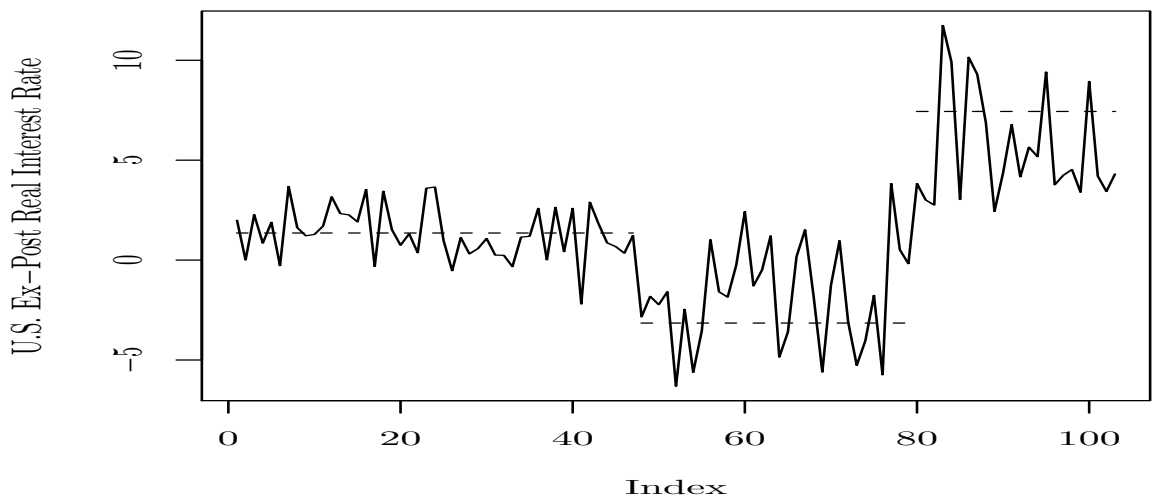}
  \caption{U.S. Ex-Post Real Interest Rate, the first quarter of 1961 -- the third quarter of 1986}
           \label{fig4}
\end{figure}

First, we need to select a $p_n$. Following the recommendations in Subsection 4.3, we will choose an
optimal $p_n$ from the range 3 to 13. For each $p_n \in \{3, 4, \ldots, 13\}$, we obtain
$\breve{\mtheta}_n$ by the ALMCPDA, and calculate the corresponding $RSS(p_n)$. Choose
$\arg\min_{3\leqslant p_n\leqslant 13}RSS(p_n)$ as the optimal $p_n$, which is 5. See Figure 5.

Based on the first step, we set $p_n=5$ and apply the five
algorithms given in Section 3 to the data. Two change points are
found based on the ALMCPDA and the CALMCPDA, which are located at 47
and 79 (see Figure 4) with RSS=455.95 corresponding to the third
quarter of 1972 and the third quarter of 1980. These results are
consistent with those of Garcia and Perron (1996). However the other
three algorithms LSMCPDA, SMCPDA and MMCPDA only detect one change
point located at 47 with RSS=1214.89. By comparing their RSSs, it is
clear that both ALMCPDA and CALMCPDA have better performance than
the other three algorithms.

\begin{figure}[!ht]
  \centering
    \includegraphics[width=1\textwidth,height=0.38\textheight]{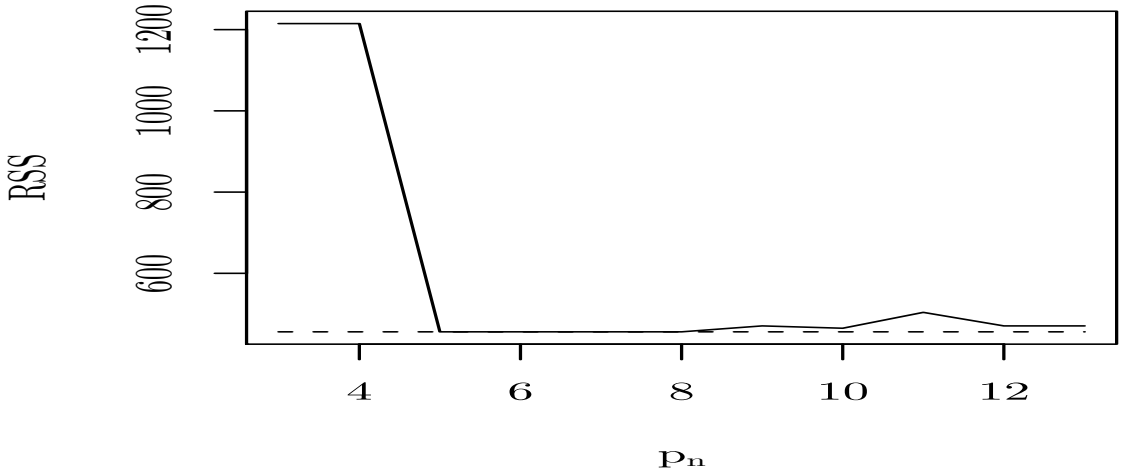}
  \caption{$RSS(p_n)$ against $p_n$ for the U.S. ex-post real interest rate data}
           \label{fig5}
\end{figure}

\vspace{.2in} \noindent{\textsf{5.2. Gross domestic product in U.S.A}}

\noindent The data presented in Maddala (1977, Table 10.3) gives the gross domestic product ($G$), the
labor input index ($L$) and the capital input index ($C$) in the United States for the years
1929-1967. $\log G$ is modeled as a linear function of $\log L$ and $\log C$. The $\log G$, $\log L$
and $\log C$ are plotted over time given in Figure 6. Worsley (1983) used the likelihood ratio method
to search for change points in this data set and pointed out that the data contained two change points
located at 1942 and 1946 (RSS$=0.011$). Caussinus and Lyazrhi (1997) used Bayes invariant optimal
multi-decision procedure to detect change points in the data series and claimed three change points
located at 1938, 1944 and 1948 (RSS$=0.01$).

\begin{figure}[!ht]
 \centering
    \includegraphics[width=1\textwidth,height=0.38\textheight]{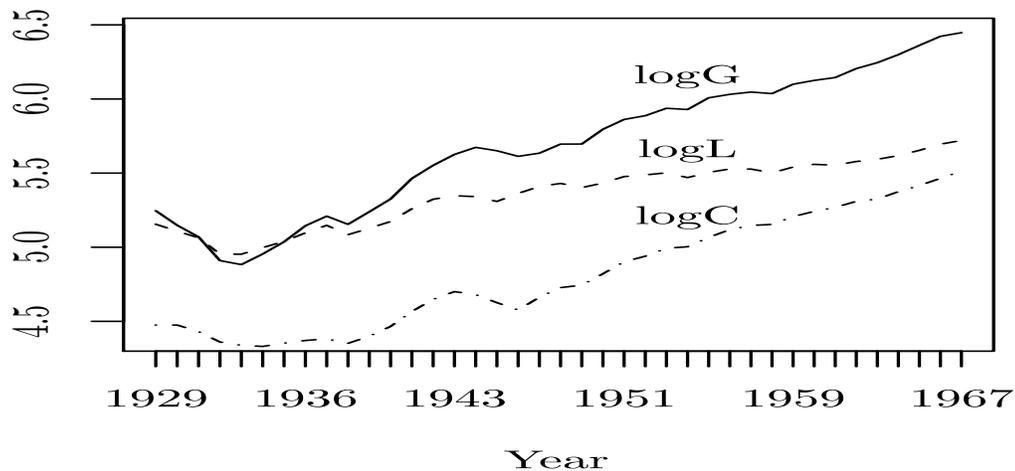}
  \caption{Logrithms of Gross domestic product ($\log$G), labor-input index ($\log$L) and capital-input index ($\log$C) in U.S.A. for the years 1929-1967.}
           \label{fig6}
\end{figure}

Since the sample size is only 39, the proposed algorithms employing
least squares or the CUSUM test may not work. Thus we only apply the
first two steps of the SMCPDA or the MMCPDA to carry out multiple
change point analysis. As in the previous example, we need to select
a $p_n$. Following the recommendations in Subsection 4.3, we will
choose an optimal $p_n$ from 13 to 17. For each $p_n\in
\{13,\ldots,17\}$, we obtain $\hat{\mtheta}^\SCAD_n$ by the SMCPDA,
and calculate the corresponding $RSS(p_n)$. Choose
$\arg\min_{13\leqslant p_n\leqslant 17}RSS(p_n)$ as the optimal
$p_n$, which is 17. With $p_n=17$, four change points detected by
applying the SMCPDA are located at 1936, 1942, 1946 and 1950 with
RSS=0.0054. With the same $p_n$, two change points detected by
applying the MMCPDA are located at 1942 and 1958 with RSS=0.015.
Thus, in terms of the RSSs,  the SMCPDA has a better performance.

\vspace{.2in} \noindent{\textbf{6. Conclusion}}

\noindent By properly segmenting the data sequence, we proposed five multiple change point detection
algorithms. The proposed approach is based on the following reasons. On the one hand, a proper
segmentation can isolate the finite change points such that each change point is only located in one
segment, and a connection between multiple change point detection and variable selection can be
established. Thus the recent advances in consistent variable selection methods such as SCAD, adaptive
LASSO and MCP can be used to detect these change points simultaneously. On the other hand, a refining
procedure using a method such as CUSUM can improve the accuracy of change point estimates. Compared
with other change point detection methods, which is very time consuming, the newly proposed algorithms
are much faster, more effective, and have strong theoretical backup. The proposed approach can be
extended to detect multiple change points in other models such as generalized linear models and
nonparametric models without any extra difficulties.

\noindent{\textbf{Appendix}}

\setcounter {equation}{0}
\def\theequation{A.\arabic{equation}}
\noindent\textsf{A.1. CUSUM test for a single change point}

\noindent Consider the following model \be \my_i=\mx_i^T\mbeta_1I(n_\ell\leqslant i \leqslant
k)+\mx_i^T\mbeta_2I(k< i \leqslant n_{\ell+1})+\varepsilon_i,\quad n_\ell\leqslant i\leqslant
n_{\ell+1},\label{ap1}\ee where $\my_{\ell}=(y_{n_\ell},\ldots,y_{n_{\ell+1}})^T$,
$\mx_{n_\ell},\mx_{n_\ell+1},\cdots,\mx_{n_{\ell+1}}$ are $q$-dimensional predictors, $\mbeta_1$ and
$\mbeta_2$ are unknown $q$-dimensional vectors of regression coefficients, and
$\me_n=(\varepsilon_{n_\ell},\ldots,\varepsilon_{n_{\ell+1}})^T$. If $n_\ell\leqslant k< n_{\ell+1}$
and $\mbeta_1\not=\mbeta_2$, there is a change point at $k$.

Let $N_\ell=n_{\ell+1}-n_{\ell}+1$. Define
\[\hat{\sigma}^2_{\ell,k}=\frac{1}{N_\ell}\left[\min_\mbeta\sum_{i=n_\ell}^k(y_i-\mx_i^T\mbeta)^2+
\min_\mbeta\sum_{i=k+1}^{n_{\ell+1}}(y_i-\mx_i^T\mbeta)^2\right],\]
and
$\hat{\sigma}^2_{\ell}=\min_\mbeta\sum_{i=n_\ell}^{n_{\ell+1}}(y_i-\mx_i^T\mbeta)^2/N_\ell$.
By Theorem 3.1.1 of Cs\"{o}rg\H{o} and Horvath (1997)), it follows
that
\begin{eqnarray}
\lim_{N_\ell\rightarrow\infty}P\left[a_\ell
\Lambda_\ell^{1/2}\leqslant
x/2+b_{\ell,q}\right]=\exp\left(-2e^{-x/2}\right),\label{cso}
\end{eqnarray}
for all $x$, where $a_\ell=(2\log \log N_\ell)^{1/2}$,
$b_{\ell,q}=2\log\log N_\ell+{q}(\log\log\log
N_\ell)/2-\log\Gamma(q/2)$, $\Gamma(x)$ is the Gamma function,
$\Lambda_\ell=\max_{n_{\ell}+q\leqslant k\leqslant
n_{\ell+1}-q}\left[-2\log\left(\hat{\sigma}^2_{\ell,k}/\hat{\sigma}^2_{\ell}\right)^{N_\ell/2}\right]$.

In light of the proof of Corollary 2.1 of Hu\v{s}kov\'a,
Pr\'a\v{s}kov\'a and Steinebach (2007), it can be shown that
\[
\lim_{N_\ell\rightarrow\infty}P\left[a_\ell
\Lambda_\ell^{1/2}\leqslant
x/2+b_{\ell,q}\right]=\lim_{N_\ell\rightarrow\infty}P\left[(\Lambda_\ell-\tilde{b}_{\ell,q})/\tilde{a}_{\ell,q}
\leqslant x\right],
\]
where $\tilde{b}_{\ell,q}=(b_{\ell,q}/a_\ell)^2$ and
$\tilde{a}_\ell={b}_{\ell,q}/a^2_\ell$, which jointly with
(\ref{cso}) implies that
\begin{eqnarray}
\lim_{N_\ell\rightarrow\infty}P\left[(\Lambda_\ell-\tilde{b}_{\ell,q})/\tilde{a}_{\ell,q}\leqslant
x\right]=\exp\left(-2e^{-x/2}\right).\nonumber
\end{eqnarray}
By Lemma 3.1.9 of Cs\"{o}rg\H{o} and Horvath (1997), it can be shown
that
\begin{eqnarray}
\lim_{N_\ell\rightarrow\infty}P\left[\left(\frac{1}{\hat{\sigma}^2_\ell}\max_{n_{\ell}+q\leqslant
k\leqslant
n_{\ell+1}-q}N_\ell(\hat{\sigma}^2_\ell-\hat{\sigma}^2_{\ell,k})-\tilde{b}_{\ell,q}\right)/\tilde{a}_{\ell,q}\leqslant
x\right]=\exp\left(-2e^{-x/2}\right).\nonumber
\end{eqnarray}
Let $T_{\ell,k}=N_\ell(\hat{\sigma}^2_\ell-\hat{\sigma}^2_{\ell,k})$
and $T_{\ell}=\max_{n_{\ell}+q\leqslant k\leqslant
n_{\ell+1}-q}T_{\ell,k}$. Given a significant level $\alpha$, the
CUSUM test for testing if there is a change point in the model
(\ref{ap1}) is given in the following: If
\begin{eqnarray}
T_\ell>\left[\tilde{b}_{\ell,q}+2\tilde{a}_{\ell,q}\log(-2/\log(1-\alpha))\right]\hat{\sigma}^2_\ell,\nonumber
\end{eqnarray}
there exists a $k\in\{n_{\ell}+q,\ldots, n_{\ell+1}-q\}$ such that
$\mbeta_1\not=\mbeta_2$ in the model (\ref{ap1}).

Denote $C_{\ell}=\sum_{i=n_\ell}^{n_{\ell+1}}\mx_i\mx_i^T$,
$\hat{\mbeta}_\ell=C_{\ell}^{-1}\sum_{i=n_\ell}^{n_{\ell+1}}\mx_iy_i$,
$C_{\ell,k}=\sum_{i=n_\ell}^k\mx_i\mx_i^T$,
$C_{\ell,k}^0=C_{\ell}-C_{\ell,k}$,
 $S_{\ell,k}=\sum_{i=n_\ell}^k\mx_i(y_i-\mx_i^T\hat{\mbeta}_\ell)$ for $k=n_\ell+q,\ldots,n_{\ell+1}-q$.
By Hu\v{s}kov\'a, Pr\'a\v{s}kov\'a and Steinebach (2007),
\begin{eqnarray}
T_{\ell}=\max_{n_{\ell}+q\leqslant k\leqslant
n_{\ell+1}-q}S_{\ell,k}^TC_{\ell,k}^{-1}C_\ell(C_{\ell,k}^0)^{-1}
S_{\ell,k}.\label{hus2}
\end{eqnarray}
Since $S_{\ell,k}$ and $C_{\ell,k}$ can be computed recursively, the
computing time of $T_\ell$ is reduced to $O(n_{\ell+1}-n_\ell)$ from
$O((n_{\ell+1}-n_\ell)^2)$ by using (\ref{hus2}).

\vspace{.2in} \noindent \textsf{A.2. Proof of Lemma 1}

\noindent Denote the elements of $\ma_c$ by $\ma_c=\{r_1, r_2,\ldots, r_{K_0}\}$. In view of
$|n-p_nm+r_km-a_{k,n}|\leqslant m$, $a_{k,n}/n\rightarrow\tau_i$, for $k=1,\ldots,K_0$ and $m=o(n)$,
by Assumption C1, it follows that \[\frac{1}{n}\sum_{i=r_1}^{r_2}
\mX_{(i)}^T\mX_{(i)}\rightarrow(\tau_2-\tau_1)W,\quad\ldots,\quad\frac{1}{n}\sum_{i=r_{K_0}}^{p_n+1}
\mX_{(i)}^T\mX_{(i)}\rightarrow(1-\tau_{K_0})W.\] Hence,
\begin{eqnarray}
&&\frac{1}{n}\mX_{\ma_c}^T\mX_{\ma_c}=U^T\begin{pmatrix}
\frac{1}{n}\sum_{i=r_1}^{r_2-1}\mX_{(i)}^T\mX_{(i)}&0&\cdots&0\\
0&
\frac{1}{n}\sum_{i=r_2}^{r_3-1}\mX_{(i)}^T\mX_{(i)}&\cdots& 0\\
 \vdots&
 \vdots&\ddots&  \vdots
\\
0& 0&\cdots& \frac{1}{n}\sum_{i=r_K}^{p_n+1}\mX_{(i)}^T\mX_{(i)}
\end{pmatrix}U\nonumber\\ \nonumber\\
&&\rightarrow\ U^T\begin{pmatrix}
(\tau_2-\tau_1)W&0&\cdots&0\\
0&
(\tau_3-\tau_2)W&\cdots& 0\\
 \vdots&
 \vdots&\ddots&  \vdots
\\
0& 0&\cdots& (1-\tau_K)W
\end{pmatrix}U\ \hat{=}\ \mathcal{W}_{\ma_c}>0,\label{wc}
\end{eqnarray}
where
\[U=\begin{pmatrix} I_{q}& 0&\cdots& 0\\
 I_{q}&
I_{q}&\cdots& 0\\
 \vdots&
 \vdots&\ddots&  \vdots
\\
I_{q}&I_{q}&\cdots& I_{q}\\
\end{pmatrix}.\]

\vspace{.2in} \noindent \textsf{A.3. Proof of Lemma 2}

\noindent As in the proof of Lemma 1, denote $\ma_c=\{r_1,\ldots,r_{K_0}\}$. It is easy to see that
\[
\tilde{X}_{n}^T\mx_\omega/m
=\frac{1}{m}\sum_{\ell=1}^{p_n}\left( \begin{array}{ll} \sum_{i=1}^n &\mx_i\mx_i^T\bomega_\ell(i)\\
\sum_{i=n-p_nm+1}^n &\mx_i\mx_i^T\bomega_\ell(i)\\
\sum_{i=n-(p_n-1)m+1}^n &\mx_i\mx_i^T\bomega_\ell(i)\\
&\vdots \\
\sum_{i=n-m+1}^n &\mx_i\mx_i^T\bomega_\ell(i)\end{array} \right).
\]
Consider the first row of $\tilde{X}_{n\ma_c}^T\mx_\omega/m$. By
Assumption C1,
$\sum_{i=n-(p_n-r_j+1)m+1}^{n-(p_n-r_j)m}\mx_i\mx_i^T/m\rightarrow
W$. Hence For large $n$,
\begin{eqnarray}
&&\left\|\frac{1}{m}\sum_{\ell=1}^{p_n}\sum_{i=1}^n\mx_i\mx_i^T\bomega_\ell(i)\right\|
\leqslant\frac{1}{m}\sum_{j=1}^{K_0}\left\|\sum_{i=n-(p_n-r_j+1)m+1}^{a_{j,n}}\mx_i\mx_i^T\mdelta_j\right\|\nonumber\\
&\leqslant&\frac{1}{m}\sum_{j=1}^{K_0}\|\mdelta_j\|\left\|\sum_{i=n-(p_n-r_j+1)m+1}^{n-(p_n-r_j)m}\mx_i\mx_i^T\right\|
\leqslant 2K_0\|W\|\max_{1\leqslant i\leqslant K_0}
\left\|\mdelta_{j}\right\|\label{l21}
\end{eqnarray}
Similarly, it can be shown that for large $n$ and $1\leqslant
s\leqslant n$,
\bea\left\|\frac{1}{m}\sum_{\ell=1}^{p_n}\sum_{i=s}^n\mx_i\mx_i^T\bomega_\ell(i)\right\|
&\leqslant&2\|W\|\left\{\begin{array}{ll}
\sum_{j=1}^{K_0}\left\|\mdelta_j\right\|,& 1\leqslant s\leqslant a_{1,n},\\
\sum_{j=2}^{K_0}\left\|\mdelta_j\right\|,& a_{1,n}< s\leqslant a_{2,n},\\
\vdots &\vdots\\
\left\|\mdelta_{K_0}\right\|,& a_{K_0-1,n}< s\leqslant a_{K_0,n},\\
0,& \mbox{elsewhere};
\end{array}\right.\nonumber\\ \nonumber \\
& \leqslant &2K_0\|W\|\max_{1\leqslant i\leqslant K_0}
\left\|\mdelta_{j}\right\|.\label{l22}\eea In view of
(\ref{l21})-(\ref{l22}),  each element of
$\tilde{X}^T_{n}\mx_\omega/m$ is bounded by
$2K_0\|W\|\max_{1\leqslant i<K_0}\left\|\mdelta_{j}\right\|$.  The
proof is complete.

\vspace{.2in} \noindent \textsf{A.4. Proof of Lemma 3}

\noindent By the definition of $\tilde{X}_{n}$, it follows that
\begin{eqnarray}
\tilde{X}_{n}^T\me_n/\sqrt{n}
=\frac{1}{\sqrt{n}} \left( \begin{array}{ll} \sum_{i=1}^n &\mx_i\varepsilon_i\\
\sum_{i=n-p_nm+1}^n &\mx_i\varepsilon_i\\
\sum_{i=n-(p_n-1)m+1}^n &\mx_i\varepsilon_i\\
&\vdots \\
\sum_{i=n-m+1}^n &\mx_i\varepsilon_i\end{array} \right).\nonumber
\end{eqnarray}
Consider the first element of $\tilde{X}_{n}^T\me_n/\sqrt{n}$. By
Assumption C1, for $j=1,\ldots,q$,
$\sum_{i=1}^nx^2_{i,j}/n\rightarrow W_{jj}$. By applying Markov's
inequality, we have
$\sum_{i=1}^nx_{i,j}\varepsilon_i/\sqrt{n}=O_p(1)$.

In the following, we show that for any $\epsilon>0$, there exists an
$M_\epsilon$ such that
\[p_{n,j}\hat{=}P\left(\frac{1}{\sqrt{n}}\max_{1\leqslant k\leqslant p_n}
\left|\sum_{i=n-(p_n-k+1)m+1}^{n}x_{i,j}\varepsilon_i\right|>M_\epsilon\right)< \epsilon.\]
Denote
$\eta_{\ell,j}=\sum_{n-(p_n-\ell+1)m+1}^{n-(p_n-\ell)m}x_{i,j}\varepsilon_i$.
Then we have
\[
p_{n,j}=P\left(\frac{1}{\sqrt{n}}\max_{1\leqslant t\leqslant
p_n}\left|\sum_{\ell=1}^{t}\eta_{p_n-\ell+1,j}\right|>M_\epsilon\right).
\]
Note that for any $v>u>0$, by Assumption C1, we have
\[Var\left(\sum_{\ell=u}^v\eta_{\ell,j}/\sqrt{m}\right)\leqslant
2(v-u)W_{jj}\sigma^2\leqslant 2(v-u)\sigma^2\max_{1\leqslant
j\leqslant q}W_{jj},\] when $n$ is large enough. By Lemma 2.1 of
Lavielle (1999), it follows that
\begin{eqnarray}
p_{n,j}&=&P\left( \max_{1\leqslant t\leqslant
p_n}\left|\sum_{\ell=1}^{t}\eta_{p_n-\ell+1,j}/\sqrt{m}\right|>M_\epsilon\sqrt{n/m}\right)\nonumber\\
 &\leqslant& \frac{cp_n}{M^2_\epsilon n/m} \leqslant
c/M^2_\epsilon<\epsilon,\nonumber
\end{eqnarray}
which means that each element of  vector
$\tilde{X}_{n}^T\me_n/\sqrt{n}$ is bounded uniformly in probability.
The proof of Lemma 3 is complete.

\vspace{.2in} \noindent \textsf{A.5. Proof of Theorem 3}

\noindent Let $\mmu=(\mmu_0^T,\mmu_1^T,\ldots,\mmu_{p_n}^T)^T$ be bounded. Put
$\mtheta=\mtheta_n+\frac{\mmu}{\sqrt{n}}$ and
\[\psi_n(\mmu)=\left|\left|\my-\mX^{(1)}_n\left(\mbeta_0+\frac{\mmu_0}{\sqrt{n}}\right)-
\sum_{j=1}^{p_n}\mX^{(j+1)}_n\left(\md_j+\frac{\mmu_j}{\sqrt{n}}\right)\right|\right|^2+\lambda_n\sum_{r=1}^{p_n}\frac{1}{|{\tilde{\md}}_{r}|^\nu}
\left|\md_{r}+\frac{\mmu_r}{\sqrt{n}}\right|.\] Let
$\breve{\mmu}_n=\arg\min\psi_n(\mmu)=\arg\min\left(\psi_n(\mmu)_n-\psi_n(\mzero)\right)$.
Thus $\breve{\mtheta}=\mtheta_n+{\breve{\mmu}_n}/{\sqrt{n}}$,  and
we only need to investigate the limiting behavior of
$\breve{\mmu}_n$. Write
$\psi_n(\mmu)-\psi_n(\mzero)\hat{=}V_n(\mmu)$, which can be
expressed as
\begin{eqnarray}
V_n(\mmu)&=&\left(\mmu^T\left(\frac{1}{n}\tilde{X}_n^T\tilde{X}_n\right)\mmu-
2\mmu^T\frac{\tilde{X}_n^T\me_n}{\sqrt{n}}-2\mmu^T\frac{\tilde{X}_n^T\mx_\omega}{\sqrt{n}}\right)\nonumber\\&+&
\frac{\lambda_n}{\sqrt{n}}\sum_{r=1}^{p_n}\frac{1}{|{\tilde{\md}}_{r}|^\nu}
\sqrt{n}\left(\left|\md_{r}+\frac{\mmu_r}{\sqrt{n}}\right|-|\md_{r}|\right).\nonumber
\end{eqnarray}
Consider the following two cases:
\begin{itemize}
\item[]Case I. For any $r\notin\ma_c$, $\mmu_r={\bf 0}$;

\item[]Case II: There are some $r\notin\ma_c$ such that $\mmu_r\neq
{\bf 0}$. Denote the number of such $r$s as $n_c$.
\end{itemize}

We first consider the case I. By Lemmas 1-2 and the assumption that
$m/\sqrt{n}\rightarrow 0$, it can be shown that as $n\to\infty$,
\begin{enumerate}
\item[(A1)] $\mmu^T\left(\frac{1}{n}\tilde{X}_n^T\tilde{X}_n\right)\mmu=\mmu_{\ma_c}^T\left(\frac{1}{n}
(\tilde{X}_{n\ma_c}^T\tilde{X}_{n\ma_c})
\right)\mmu_{\ma_c}\rightarrow\mmu_{\ma_c}^T\mathcal{W}_{\ma_c}\mmu_{\ma_c};$
\item[(A2)]  $\mmu^T\tilde{X}_n^T\me/\sqrt{n}=\mmu_{\ma_c}^T(\tilde{X}_{n\ma_c}^T\me)/\sqrt{n}\rightarrow_d
\mmu_{\ma_c}^T\mw_{\ma_c},$ where
$\mw_{\ma_c}=N(\mzero,\sigma^2\mathcal{W}_{\ma_c})$;
\item[(A3)] $\mmu^T\tilde{X}_n^T\mx_\omega/\sqrt{n}\rightarrow 0$.
\end{enumerate}
Note that for any $ r\notin\ma_c$, the second term of $V_n(\mmu)$
equals to 0. Let $r\in \ma_c$. By Assumption C3, it follows that
$1/|\tilde{\md}_r|^{\nu}\leqslant c^{-\nu}$ in probability. Since
$\sqrt{n}\left||\md_r+\frac{\mmu_r}{\sqrt{n}}|-|\md_r|\right|
\leqslant |\mmu_r|$, and $|\ma_c|=K_0$, by the assumption that
$\lambda_n/\sqrt{n}\to0$, we have
\[\frac{\lambda_n}{\sqrt{n}}\displaystyle\sum_{r=1}^{p_n}\frac{1}{|\tilde{\md}_r|^{\nu}}
\sqrt{n}\left(\left|\md_r+\frac{\mmu_r}{\sqrt{n}}\right|-|\md_r|\right)
\rightarrow_p 0,\] which, jointly with (A1)-(A3) above, implies that
$V_n(\mmu)\rightarrow_p \mmu^T_{\ma_c}
\mathcal{W}_{\ma_c}\mmu_{\ma_c}-2\mmu^T_{\ma_c}\mw_{\ma_c},$ as
$n\to\infty$.

We now consider the case II. By Lemmas 2-3 and the assumption that
$m/\sqrt{n}\rightarrow 0$, it can be shown that
\begin{enumerate}
\item[(B1)] $\mmu^T\left(\frac{1}{n}\tilde{X}_n^T\tilde{X}_n\right)\mmu\geqslant0$;
\item[(B2)]  $\mmu^T\tilde{X}_n^T\me_n/\sqrt{n}=O_p(n_c)$;
\item[(B3)] $\displaystyle\frac{1}{n_c}\mmu^T\tilde{X}_n^T\mx_\omega/\sqrt{n}\rightarrow 0$.
\end{enumerate}
As argued previously, it can also be shown that
\begin{enumerate}
\item[(B4)] $\frac{\lambda_n}{\sqrt{n}}\sum_{r\in\ma_c}\frac{1}{|{\tilde{\md}}_{r}|^\nu}
\sqrt{n}\left(\left|\md_{r}+\frac{\mmu_r}{\sqrt{n}}\right|-|\md_{r}|\right)\to0.$
\end{enumerate}
Now let $r\notin \ma_c$.  Since $\left|\{r,\md_r={\bf{0}},
\mmu_r\neq\mzero\}\right|$ $=n_c$,  by Assumption C3 and the
assumption that
$\lambda_n(n/p_n)^{\nu/2}/\sqrt{n}\rightarrow\infty$, it follows
that \beaa &&\frac{1}{n_c}\sum_{r\not\in\ma_c,
{\md_r}={\bf{0}},{\mmu_r\neq}{\bf{0}}}\frac{\lambda_n}{\sqrt{n}}\frac{1}{|\tilde{\md}_r|^{\nu}}
\sqrt{n}\left(\left|\md_r+\frac{\mmu_r}{\sqrt{n}}\right|-|\md_r|\right)\\
&=& \frac{1}{n_c}\sum_{r\not\in\ma_c,
{\md_r}={\bf{0}},{\mmu_r\neq}{\bf{0}}}\frac{\lambda_n}{\sqrt{n}}
\left(\frac{n}{p_n}\right)^{\nu/2}|\mmu_r|\times
\left|\sqrt{\frac{n}{p_n}}\tilde{\md}_r\right|^{-\nu}
\rightarrow_p\infty,\eeaa which, jointly with (B1)-(B4), implies
that $V_n(\mmu)\rightarrow_p \infty.$

So far we have showed that \be V_n(\mmu)\rightarrow_p V(\mmu)=
\begin{cases} \mmu^T_{\ma_c}\mathcal{W}_{\ma_c}\mmu_{\ma_c}-
2\mmu^T_{\ma_c}\mw_{\ma_c}, &\text{Case I},\\
\infty, &\text{Case II}.
\end{cases}\label{convex}\ee
It can be seen that $V$ is a convex function and has a unique
minimum at $\breve{\mmu}$ such that
$\breve{\mmu}_{\bar{\ma}_c}=\mzero$ and
$\breve{\mmu}_{\ma_c}=\mathcal{W}_{\ma_c}^{-1}\mw_{\ma_c}$. Since
$V_n(\cdot)$ is also a convex function and has a unique minimum
denoted by $\breve {\mmu}_n$, by (\ref{convex}),
\[\breve {\mmu}_n=\arg\min V_n(\mmu)\rightarrow_p\arg\min V(\mmu)=\breve{\mmu},\]
and hence,
\[(\breve {\mmu}_n)_{\ma_c}\rightarrow_p
\breve{\mmu}_{\ma_c}=\mathcal{W}_{\ma_c}^{-1}\mw_{\ma_c}\quad
\text{and}\quad (\breve
{\mmu}_n)_{\bar{\ma}_c}\rightarrow_p\breve{\mmu}_{\bar{\ma}_c}=\mzero.\]
In view of the fact that $\mw_{\ma_c}\sim
N(\mzero,\sigma^2\mathcal{W}_{\ma_c})$, the proof is complete.

\vspace*{8pt}

\noindent{\textbf{Acknowledgements}}

This work was supported by the Natural Sciences and Engineering Research Council of Canada. The
authors thank Professor Pierre Perron for his kindly sharing the U.S. ex-post real interest rate data
with them.

\vspace{.2in} \noindent{\textbf{References}} \vskip 5mm

\begin{description}

\item Chen, J., and Gupta, A.K. (2000). Parametric Statistical Change Point
Analysis, \emph{Birkh\'auser}.

\item Cs\"{o}rg\H{o},\ M, Horvath, L., (1997). Limit Theorems in Change-Point
Analysis, \emph{Chichester:Wiley}.

\item Davis, R.A., Huang, D., and Yao Y.C. (1995). Testing for a Change in the Parameter
Values and Order of an Autoregressive Model, \emph{The Annals of Statistics}, {23}, 282-304.

\item Davis, R.A., Lee, T.C.M., and Rodriguez-Yam, G.A. (2006). Structural Break Estimation for Nonstationary
Time Series Models, \emph{Journal of the American Statistical Association}, {101}, 223-239.

\item Efron, B., Hastie, T., Johnstone, I. and Tibshirani, R. (2004). Least Angle Regression, \emph{Annals of Statistics}, {32}, 407-499.

\item Fan, J., and Li, R. (2001). Variable Selection via Nonconcave Penalized Likelihood and Its
Oracle Properties, \emph{Journal of the American Statistical Association}, {96}, 1348-C1360.

\item Fan, J., and Lv, J. (2008). Sure Independence Screening for Ultrahigh Dimensional Feature Space,
\emph{Journal of the Royal Statistical Scociety}, Ser. B, {70}, 849-911.

\item Garcia, R., and Perron, P. (1996). An Analysis of the Real Interest Rate under Regime Shifts, \emph{The Review of Economics and Statistics}, {78}, 111-125.

\item Harchaoui, Z., and Levy-Leduc, C. (2008). Catching Change-Points with Lasso, \emph{Advances in Neural Information Processing Systems}.

\item Huang J., Ma, S., and Zhang, C. (2008). Adaptive Lasso for Sparse High-dimensional Regression Models,
\emph{Statistica Sinica}, {18}, 1603-1618.

\item Hu\v{s}kov\'a, M., Pr\'a\v{s}kov\'a, Z., and Steinebach, J. (2007). On the detection of
changes in autoregressive times series I. Asymptotics, \emph{Journal of Statistical Planning and
Inference}, {137}, 1243-1259.

\item Kim, H.-J., Yu, B., and Feuer, E.J. (2009). Selecting the
Number of Change-Points in Segmented Line Regression, \emph{Statistica Sinica}, {19}, 597-609.

\item Kuelbs, J., and Philipp, W. (1980). Almost Sure Invariance Principles for
Partial Sums of Mixing $B$-Valued Random Variables, \emph{The Annals of Probability}, {8}, 1003-1036.

\item Lavielle, M. (1999). Detection of Multiple Changes in a Sequence of Dependent
Variables, \emph{Stochastic Processes and their Applications}, {83}, 79-102.

\item Loschi, R.H., Pontel, J.G., and Cruz, F.R.B. (2010). Multiple
Change-Point Analysis for Linear Regression Models, \emph{Chilean Journal of Statistics}, 1, 93-112.

\item Pan, J., and Chen, J. (2006). Application of Modified Information
Criterion to Multiple Change Point Problems, \emph{Journal of Multivariate Analysis}, {97}, 2221-2241.

\item Tibshirani, R. (1996). Regression Shrinkage and Selection via the Lasso, \emph{Journal of the Royal Statistical Society}, Ser. B, {58}, 267-288.

\item Wang, H., Li, G., and Tsai, C. (2007). Regression Coefficient and Autoregressive Order
Shrinkage and Selection via the Lasso, \emph{Journal of the Royal Statistical Scociety}, Ser. B, {69},
63-78.

\item Zhang, C., and Huang, J. (2008). The Sparsity and Bias of the Lasso Selection in
High-dimensional Linear Regression, \emph{The Annals of Statistics}, {36}, 1567-1594.

\item Zhang, C. (2010). Nearly Unbiased Variable Selection Under Minimax Concave Penalty,
\emph{The Annals of Statistics}, {38}, 894-942.

\item Zhao, P., and Yu, B. (2006). On Model Selection Consistency of
Lasso, \emph{Journal of Machine Learning Research}, {7}, 2541-2567.

\item Zou, H. (2006). The Adaptive Lasso and Its Oracle Properties, \emph{Journal of the American Statistical Association}, {101}, 1418-1429.

\end{description}
\end{document}